\documentclass[pdflatex,sn-mathphys-num]{sn-jnl}

\usepackage{graphicx}%
\usepackage{multirow}%
\usepackage{amsmath,amssymb,amsfonts}%
\usepackage{amsthm}%
\usepackage{mathrsfs}%
\usepackage[title]{appendix}%
\usepackage{xcolor}%
\usepackage{textcomp}%
\usepackage{manyfoot}%
\usepackage{booktabs}%
\usepackage{algorithm}%
\usepackage{algorithmicx}%
\usepackage{algpseudocode}%
\usepackage{listings}%
\usepackage{titletoc}
\usepackage{forest}
\usepackage{bbm}
\usepackage{xcolor}
\usepackage[T1]{fontenc}

\theoremstyle{thmstyleone}%
\theoremstyle{thmstyletwo}%

\theoremstyle{thmstylethree}%

\begin{document}

\title[Why Human Guidance Matters in Collaborative Vibe Coding]{Why Human Guidance Matters in Collaborative Vibe Coding}
\author[1]{\mbox{Haoyu Hu}}
\author[2]{ \mbox{Raja Marjieh}}
\author[2,3,4]{\mbox{Katherine M Collins}}
\author[1]{\mbox{Chenyi Li}}
\author[2]{\mbox{Thomas L. Griffiths}}
\author[5]{\mbox{Ilia Sucholutsky}\textsuperscript{\textdagger}}
\author[1]{\mbox{Nori Jacoby}\textsuperscript{\textdagger}}
\affil[1]{Cornell University}
\affil[2]{Princeton University}
\affil[3]{Massachusetts Institute of Technology}
\affil[4]{University of Cambridge}
\affil[5]{New York University}
\affil[]{\textsuperscript{\textdagger}Joint senior author}

\maketitle
\begin{abstract}

Writing code has been one of the most transformative ways for human societies to translate abstract ideas into tangible technologies. Modern AI is changing this process by enabling experts and non-experts alike to generate code without actually writing it, instead using natural language instructions or ``vibe coding''. While increasingly popular, the  impact of vibe coding on productivity and collaboration, and the role of humans in this process, remains unclear. Here, we introduce a controlled experimental framework for studying collaborative vibe coding and use it to compare human-led, AI-led, and hybrid groups. Across 20 experiments involving 737 human participants, we show that people provide uniquely effective high-level instructions for vibe coding, whereas AI-provided instructions often result in performance collapse. We further demonstrate that hybrid systems perform best when humans lead by providing instructions while evaluation is delegated to AI. Although AI systems can rapidly optimize performance for specific tasks, our work highlights the importance of human guidance in shaping future hybrid societies.

\end{abstract}

\section{Introduction}

Coding is a foundational skill in contemporary digital societies, supporting a wide range of economic, scientific, and creative activities \citep{kelly2023starting,hu2024programming}. In recent years, advances in large language models have led to rapid growth in AI-assisted programming tools, accompanied by major investments from companies \citep{wong2023natural,jiang2024survey}. Systems such as Copilot, Cursor, and related tools are now widely used in real-world development workflows \citep{wermelinger2023using,kumar2025evaluating}. These tools do more than speed up coding and may even act as collaborative ``thought partners'' \citep{collins2024building}. This has the potential to not only change the kind of code people write, but how they approach coding as a practice.




One new practice in human-AI collaborative coding is \textit{vibe coding}. Vibe coding describes a conversational and intuition-driven style of programming whereby users provide high-level guidance and rely on iterative interaction with AI systems, rather than having the traditional precise low-level control over their code \citep{ray2025review,sapkota2025vibe,fawzy2025vibe,geng2025exploring,collins2025ai}. Instead of writing code, users describe goals, intentions, or desired changes, while AI systems generate and modify code accordingly. This interaction style reframes coding as a collaborative process between humans and AI, rather than a purely human activity.

These developments point toward the emergence of ``hybrid societies'' in which humans and AI agents jointly produce complex artifacts \citep{shiiku2025dynamics,hoche2025ai}. In such systems, performance depends not only on the capabilities of individual AI models, but also on how humans and AI agents coordinate over time, how responsibilities are divided, and how guidance, evaluation, and execution are distributed across agents \citep{wang2025ai,liu2024toward,sapkota2025vibe}. Despite the importance of these questions, we still have a limited empirical understanding of how AI-assisted coding reshapes collective outcomes. Most prior work focuses on improving model accuracy, interface design, or single interaction usability, rather than the iterative process of collaboration and coordination in coding \citep{li2025large,alam2024enhancing,amershi2019guidelines}.

This gap persists in part because human–AI collaboration is difficult to study in the wild at scale~\citep{sucholutsky2025using}. Many studies rely on simulations of fully autonomous AI agents or offline benchmarks \citep{tufano2024autodev,lyu2025automatic}, leaving open the question of whether direct human involvement -- and what type of human involvement -- changes outcomes in systematic ways. This raises two central questions: (i) is there a unique advantage to human–AI collaboration over fully automated AI pipelines?; and (ii) how should labor be effectively divided within groups that include both humans and AI? 


To begin to address these questions, we introduce a new controlled experimental framework for studying collaborative vibe coding. Our approach affords full causal control over who provides guidance, who executes code, who validates outputs, and how information flows across iterations. We deliberately design a task that is intuitive for non-experts, supports repeated interaction, and allows for objective validation against a known target. In this task, we systematically compare human- and AI-driven vibe coding, analyze sources of misalignment, test hybrid collaboration strategies, and examine how different role allocations affect collective performance.
Our goal is to determine under which conditions human–AI collaboration is most effective, and to assess what contribution people provide over fully automated AI coding systems.

\subsection{Experimental Approach}

\begin{figure*}
    \centering
    \includegraphics[width=\linewidth]{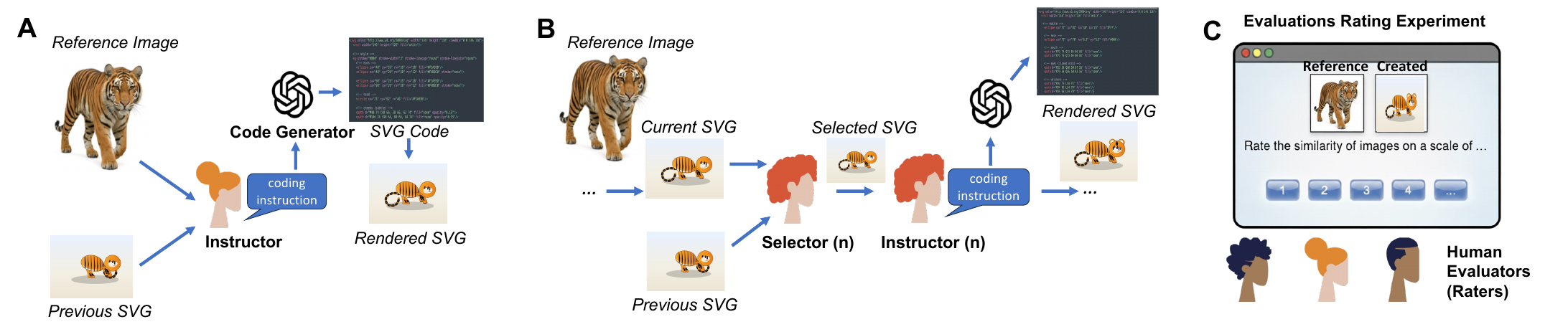}
    \caption{Vibe coding experimental paradigm. Participants generate SVG code, which can be rendered into images. (A) Core procedure. An instructor views a reference image and its best SVG rendition from the previous iteration, and uses natural language instructions to guide code-generation. The code generator produces SVG code that is then rendered into an image. (B) Iterated procedure. At each iteration, a selector chooses whether the current or the previous SVG image better matches the reference image. The selected SVG image is then passed to an instructor, who provides vibe-coding instructions that are carried forward to the next iteration. (C) Interface of the human validation experiments. Participants rate the similarity of generated SVG images to the reference image.
    }
    \label{fig:figure1}
\end{figure*}

Writing code allows people to translate abstract ideas into actions performed by machines. In this paper, we focus on one type of idea transfer: translating a visual idea (e.g., a mental image) into a graphical representation. In our setting, we provide a reference image as a stand-in for the idea people may want to realize (Fig. \ref{fig:figure1}), and the goal of the coding task is to recreate this visual idea through code. For simplicity, we use scalable vector graphics (SVG) as the coding medium. SVG is a programming language that can be directly rendered into an image, making it possible to easily judge whether the output matches the intended target. This makes SVG well-suited for studying collaborative vibe coding, as participants can interpret the code by immediately observing its visual output. We created 10 reference images by prompting GPT-5 \citep{singh2025openai} to generate images of the following animals: cat, dog, tiger, bird, elephant, penguin, shark, zebra, giraffe, panda. 

Fig. \ref{fig:figure1}A  illustrates the setup used in our experiments (see screenshots from the experiment in Appendix \ref{appendix:experiment-interface}). An \textit{instructor} (vibe coder) views a reference image and its SVG rendition from the previous iteration and then uses natural language instructions to guide a code-generation system. The \textit{code-generator} produces SVG code that can then be rendered as an image. The process is iterative, emulating real-world coding practices: each instructor starts from the best result of the previous iterations and incrementally refines it to better match the target image.


We also introduce a selection step to mimic realistic collaborative coding workflows. At each stage, the \textit{selector} compares the current version with the previous one and can revert the changes if they do not improve the result. This ability to select and discard modifications reflects common processes of technical refinement, such as those used in version control systems (e.g., Git). Fig. \ref{fig:figure1}B shows the overall iterative procedure (see screenshots  in Appendix \ref{appendix:experiment-interface}). At each iteration, a selector decides whether the current or the previous SVG better matches the reference image. The selected SVG is then passed to an instructor, who provides vibe-coding instructions that are passed forward.

In the basic experimental configuration, the same participant performed both the selection and instruction roles at each iteration. In the first iteration, no previous SVG was provided, and the instructor generated an image solely based on the reference image. Likewise, in the second iteration, no selector was involved, as only a single image had been created at that point. 


Finally, we simulate the role of independent \textit{evaluators} by involving an independent group of human participants. These evaluators view both the reference image and the generated SVG output and assess the quality of the vibe-coding process by rating the visual similarity between the two images (Fig. \ref{fig:figure1}C, Appendix \ref{appendix:validation-interface}). As an additional control, we also had AI systems perform the evaluation task. 
An extended Methods section is provided in Appendix~\ref{smethods}. All experiment codes and data are available at \href{https://doi.org/10.17605/OSF.IO/BZDRV}{https://doi.org/10.17605/OSF.IO/BZDRV}.

\section{Results}\label{results}
\subsection{Comparing human and AI roles}\label{results:human-ai main comparison}
A total of 30 independent vibe-coding ``chains'' with 15 iterations were run (three repeats for the 10 reference images). In one condition, we recruited 49 human participants serving as both selectors and instructors (\textit{human-led} vibe coding condition). Within each chain, the same participant first acted as the selector and then as the instructor. In a second condition (\textit{AI-led} vibe coding), OpenAI’s GPT-5 model performed the same task as human participants, with all other aspects of the experiment remaining identical. In all cases, instructors and selectors used natural language to prompt a code-generation system (GPT-5).
\begin{figure*}
    \centering
    \includegraphics[width=\linewidth]{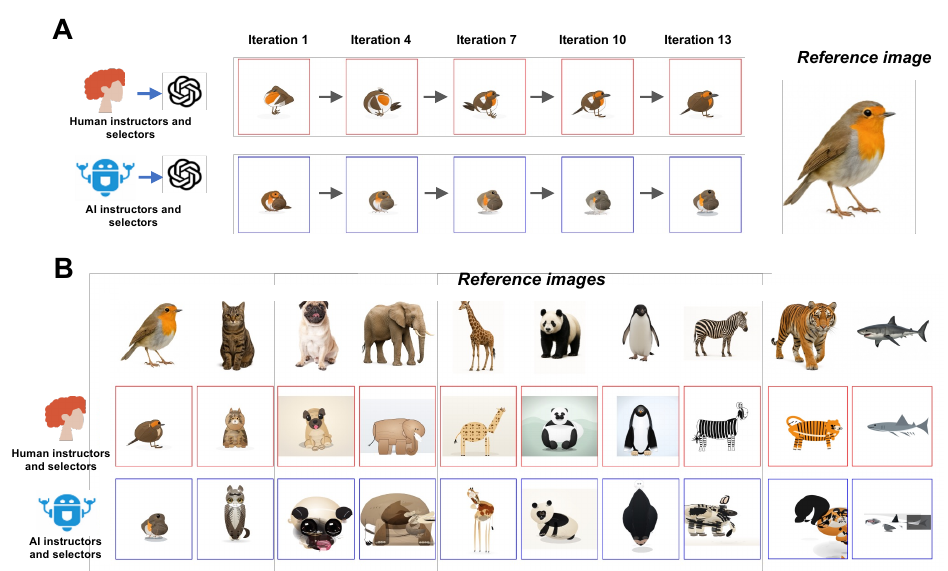}
    \caption{Example of human-led (human selectors and instructors) and AI-led (AI selectors and instructors) vibe coding. (A) Example progressions of the experiment for one reference image with human-led (top) and AI-led (bottom) vibe coding. (B) Examples from the last iteration of human-led (top) and AI-led (bottom) chains. Additional examples in Appendix \ref{appendix:more-cases}.}
    \label{fig:figure2}
\end{figure*}
Fig. \ref{fig:figure2}A shows a representative example of multi-round vibe coding for one reference image. With humans as the selectors and instructors (Fig. \ref{fig:figure2}A top), the SVG output improves steadily across iterations. In contrast, in the AI-led condition (Fig. \ref{fig:figure2}A bottom), early iterations sometimes capture salient features, but later iterations do not substantially improve or even drift away from the target.
Fig. \ref{fig:figure2}B suggests a consistent human advantage in the final SVG outputs (15th iteration) across all target images under both conditions.

\begin{figure}[t]
  \begin{center}
    \includegraphics[width=0.8\columnwidth]{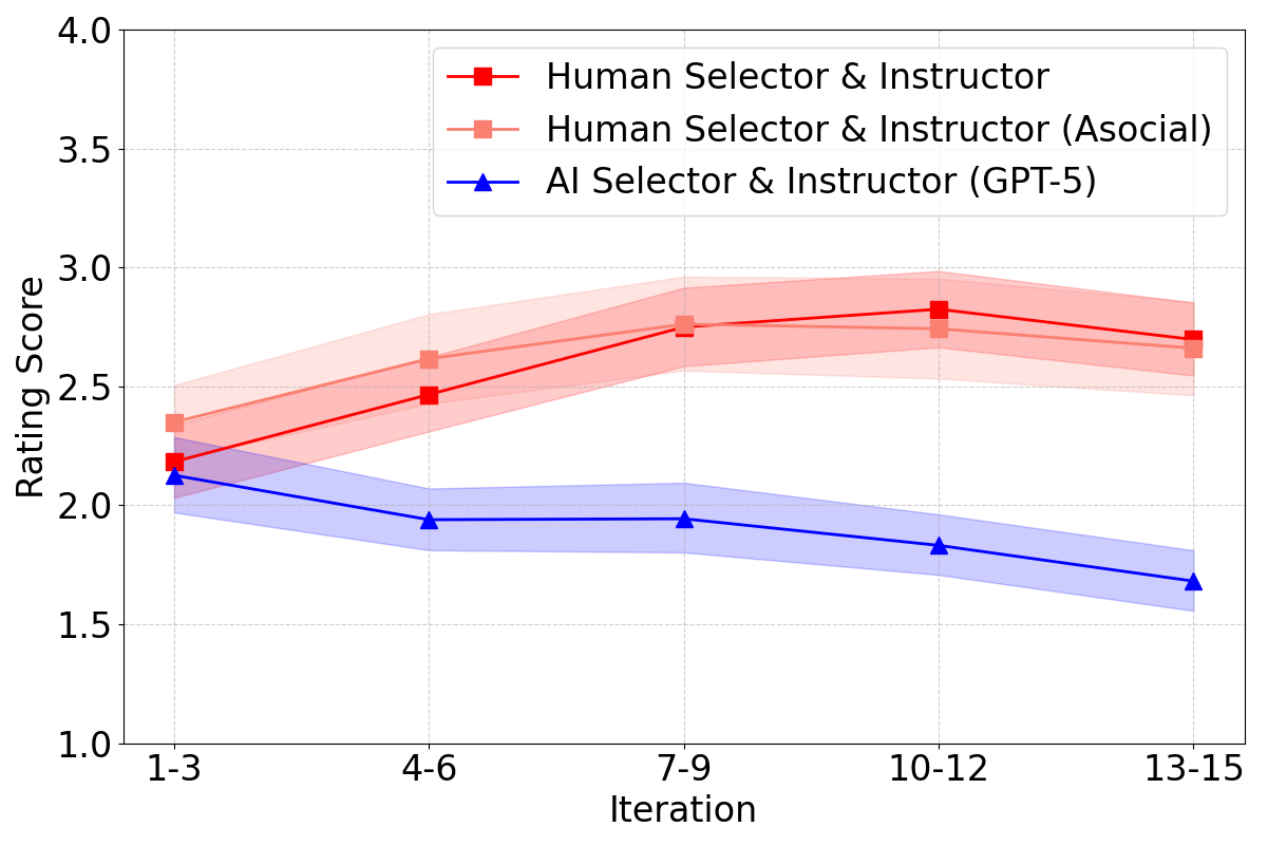}
  \end{center}
  \caption{Performance of human- and AI-led vibe coding. Data points represent average rating scores across all experiments. An additional asocial control whereby individual participants complete the full vibe-coding process (within-participant design) is also provided. Shaded area represents 95\% confidence interval.}
  \label{fig:figure2-extra}
\end{figure}
To test this quantitatively, we recruited a separate group of human evaluators to rate the resulting images based on their similarity to the reference image (Fig. \ref{fig:figure2-extra}; design described in Fig. \ref{fig:figure1}C).
In the first iterations, humans and AI systems attained similar ratings ($\Delta=.06$, $p=.304$, Cohen's $d=.08$; statistical methods described in Appendix \ref{appendix:rating-data-analysis}), however in the final iteration, humans significantly surpassed AI ($\Delta=1.01$, $p<.001$, $d=1.49$).
When human selectors and instructors were involved, similarity scores increased generally across iterations (indicated by a positive correlation between rating scores and iterations: $r=.25$, $\text{CI}_{95\%}$=[.16, .33]) with relative improvement reaching 23.4\%.
In contrast, in the AI-led baseline, similarity scores significantly decreased across iterations (indicated by a negative correlation $r=-.23$, $\text{CI}_{95\%}$=[-.31, -.13]).
These results demonstrate a clear advantage of humans acting as selectors and instructors over a fully AI-driven coding pipeline. Moreover, while humans exhibit iterative refinement, AI systems deteriorate.

As we are using a single model in the AI-led condition while involving different participants in the human-led condition, the performance difference may be attributed to the collective intelligence of human groups. To test this, we recruited another 10 participants, each independently completing 3 chains (Asocial condition; Fig. \ref{fig:figure2}C). The performance was nearly identical and increased continuously ($r=.12$, $\text{CI}_{95\%}$=[.03, .21]). 
These results rule out individual differences in vibe coding ability as the primary driver of performance gains and suggest that improvement arises from alignment with the task objective rather than from social interaction per se. In other words, performance gains reflect human-guided iterative refinement and are largely independent of whether improvements are produced by a single individual or by a group.

To examine whether the performance difference between AI and human instructors reflects a misalignment in their visual representations \citep{sucholutsky2023getting}, we conducted an additional experiment in which the evaluator role was replaced by GPT-5 (Appendix \ref{appendix:ai-rating}). Specifically, we asked whether AI agents would recognize that their own outputs are inferior to those produced by humans, or instead show a preference for their own creations, which would indicate a potential alignment issue.

Overall, AI evaluators assigned higher ratings than human evaluators across all creative outputs ($\Delta=2.083$, $p<.001$, $d=2.640$). However, AI evaluators were less sensitive in discriminating between human- and AI-led stimuli (Over all iterations, human vs. AI, as evaluated by AI: $\Delta=-.06$, $p=.899$, $d=-.08$; human vs. AI, as evaluated by humans: $\Delta=.68$, $p<.001$, $d=.91$).
Notably, in the early iterations, AI rated a higher score in AI-led vibe coding results than human-led results ($\Delta=.34$, $p<.001$, $d=.49$). This asymmetry emerged despite initially minimal baseline differences in human evaluations of human versus AI outputs, as mentioned above. Together, these findings indicate that, particularly in early iterations, AI evaluators exhibit a measurable preference for their own outputs. These findings suggest that the observed performance differences may stem from a misalignment in representations between humans and AI or from reduced overall sensitivity of AI to quality differences \citep{huang2024characterizing}.
\begin{figure*}[!htb]
    \centering    \includegraphics[width=\linewidth]{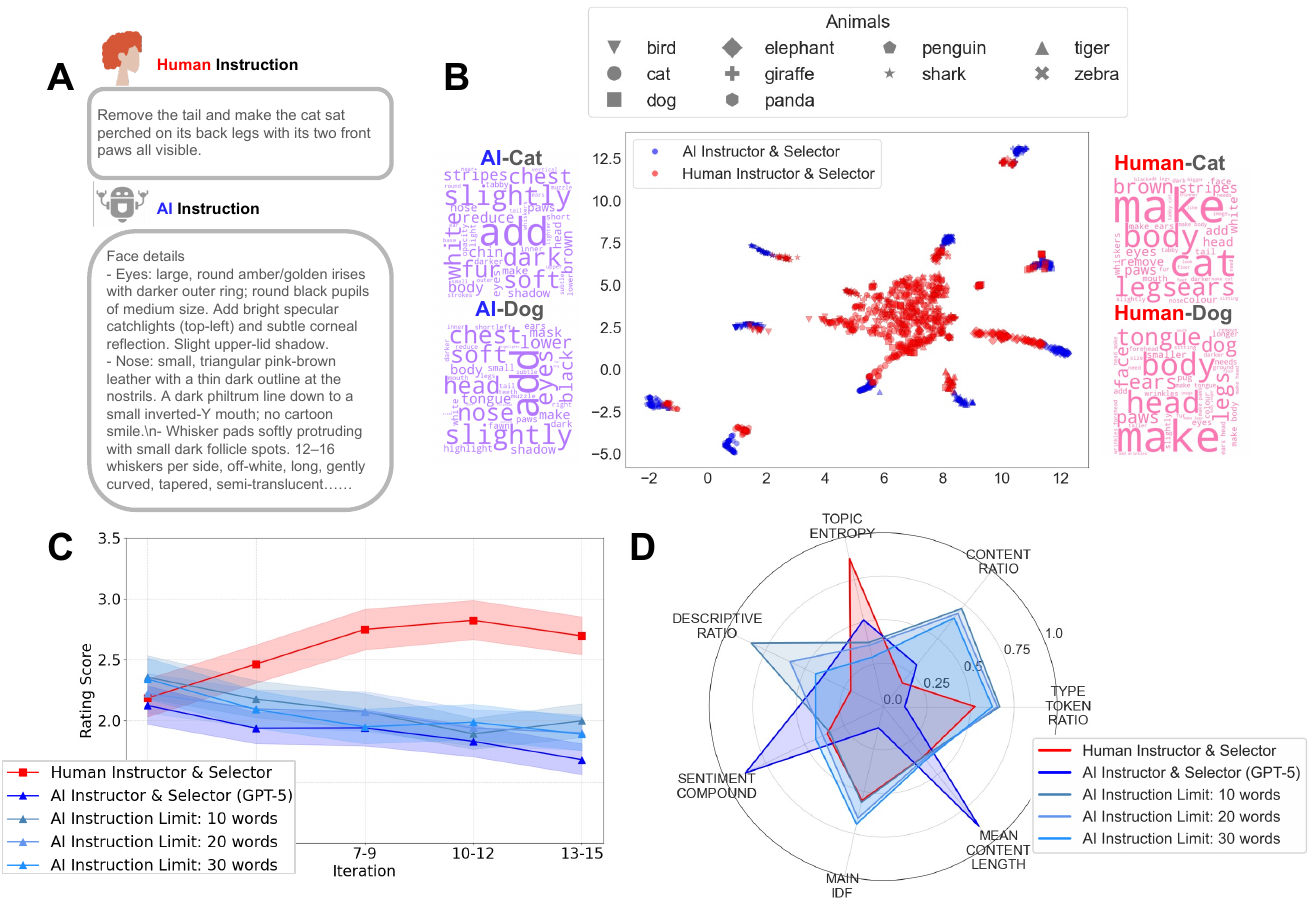}
    \caption{Comparing instruction semantics in human-led and AI-led conditions. (A) Example human- and AI-generated instructions. (B) UMAP projection of instructions in a semantic embedding space and example word clouds of both human and AI instructions. (C) Validation experiment results from different AI-led experiments under instruction length limits. (D) Radar plot of seven semantic metrics: topic entropy (the diversity and unpredictability of topics), descriptive ratio (proportion of descriptive words), sentiment compound (overall emotion of the text), main IDF (how unique the vocabulary is), mean content length (average length of the answer), type token ratio (ratio of unique words to the total answer) and content ratio (proportion of content words vs. function words). For comparison, each metrics was normalized to the scale of 0 to 1. Additional details provided in Appendix \ref{appendix:semantic-analysis}.}
    \label{fig:figure3}
\end{figure*}

\subsection{Divergences between human and AI instruction semantics}\label{results:semantic}
To further explore what drove differences in the outcomes of human and AI instructions, we next analyzed the text of the instructions. Fig. \ref{fig:figure3}A shows representative examples of human and AI instructions. Human instructions are typically short and goal-directed. In contrast, AI instructions are much longer and more complex, often containing exhaustive descriptions of visual attributes, including fine-grained details about texture, lighting, color gradients, facial features, and animal-specific anatomical properties.

To compare the semantic differences between human and AI instructions, we converted all instructions into sentence embeddings using GPT text-embedding-3-small and projected them into a two-dimensional space using UMAP (\cite{mcinnes2018umap}; Fig. \ref{fig:figure3}B). Human instructions form a single dense semantic cluster, indicating high consistency across targets and iterations. In contrast, AI instructions are split into multiple target-specific clusters. This suggests that humans reuse a shared, task-oriented language across different targets, while AI treats each target animal image as a separate descriptive problem. Fig. \ref{fig:figure3}B shows TF–IDF word clouds \citep{leskovec2020mining} demonstrating this phenomenon. Human word clouds are dominated by general action words such as ``make'', ``body'', and ``head'' which are reused across targets. AI word clouds contain more target-specific and descriptive terms, for example, using words ``soft'', ``slightly'' frequently and using ``fur'' for cat and ``tongue'' for dog.  We then computed, across animal categories, the cosine similarity with average instruction embeddings. Human instructions showed higher similarities across categories (mean similarity $M=.84$,  $\text{CI}_{95\%}$=[.83, .85]) compared with AI instructions ($M=.78$,  $\text{CI}_{95\%}$=[.76, .79]). The difference was significant ($\Delta=.06$, $p<.001$, $d=1.16$).

AI instructions were much longer than human instructions (on average number of instruction words for humans was $M=17.72$, $SD=16.17$, and for AI was $M=755.28$, $SD=193.80$, $p<.001$, $d = 6.48$). We next tested whether instruction length alone explains the performance gap. We reran the AI-led vibe coding condition with hard limits of 10, 20, or 30 words per instruction (Fig. \ref{fig:figure3}C). After applying length limits, the mean content length of AI instructions became comparable to human instructions. However, limiting instruction length did not improve AI performance, continuing to show flat or declining performance over iterations ($r=-.20$, $\text{CI}_{95\%}$=[-0.28, -0.10], $r=-.15$, $\text{CI}_{95\%}$=[-.24, -.06], $r=-.19$, $\text{CI}_{95\%}$=[-.28, -.10] for 10, 20 and 30 iterations),
indicating that verbosity alone does not account for the observed failure. Moreover, the semantic clustering of AI instructions remained substantially tighter than that of human instructions even when controlling for instruction length. For descriptions limited to 10, 20, and 30 words, AI instructions showed significantly less clustering than human instructions (within category similarity means $M=.64$, $.73$, $.76$, respectively; compared with humans $M=.40$, all differences $p<.001$), indicating that AI remains overly target-specific even when constrained to short descriptions.

To further characterize the misalignment, we calculated seven semantic metrics for all instructions (Fig. \ref{fig:figure3}D, Appendix \ref{appendix:semantic-analysis}). After applying length limits, the mean content length of AI instructions became comparable to human instructions. However, AI instructions showed a higher descriptive ratio, indicating a greater proportion of descriptive words than action-oriented content. This suggests that AI focuses on a detailed depiction of the target rather than on communicating effective guidance to the coding assistant. In addition, humans exhibited higher topic entropy than all AI conditions, indicating greater diversity and flexibility in how they approached improvements across iterations. This pattern suggests that humans are better at identifying what is missing or incorrect in the current SVG and at exploring alternative ways to fix it, whereas AI remains locked into narrow, over-specified descriptions.

\subsection{Testing hybrid human-AI instructors and selectors}\label{results:hybrid}

Building on the previous results showing a strong advantage of humans as selectors and instructors, we next asked whether this advantage could be partially retained while reducing human involvement. Human labor is costly and difficult to scale, whereas AI can provide fast and generally less expensive automatic guidance. We next tested a hybrid human–AI vibe coding setting that mixes human and AI vibe coders within the same multi-round coding chain with the same code generation system.
\begin{figure*}
  \begin{center}
    \includegraphics[width=\linewidth]{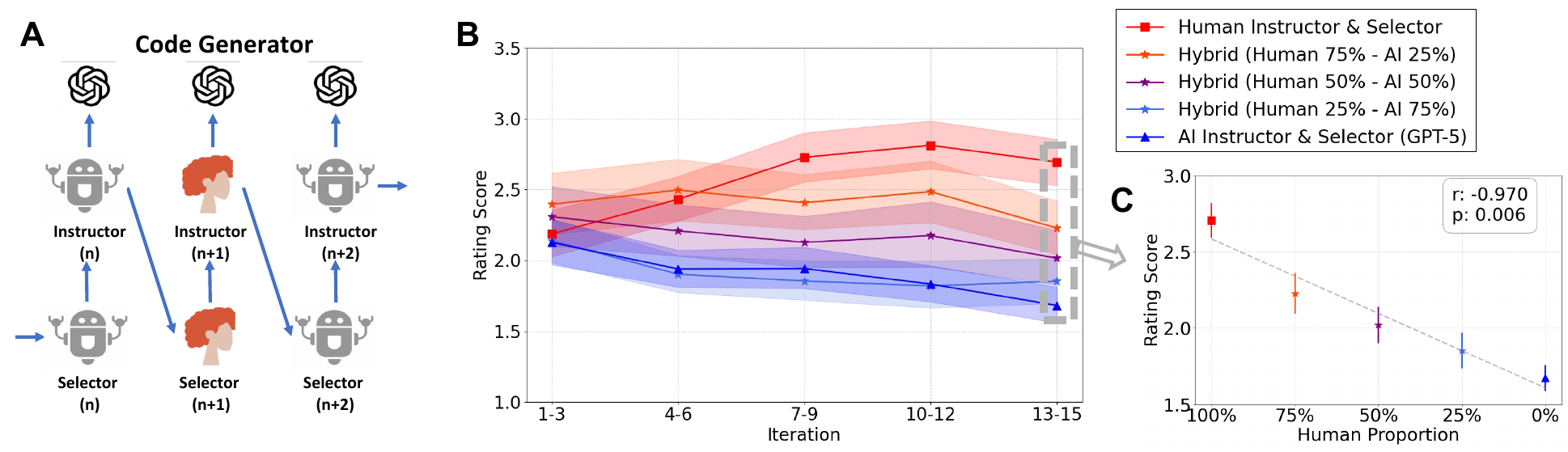}
  \end{center}
  \caption{Hybrid Human-AI-Led vibe coding. (A) The schematics of human-AI hybrid vibe coding. (B) Validation rating results of human-AI hybrid vibe coding. (C) Trade-off between human proportion and the final vibe coding performance. Error bars correspond to 95\% CI, same as the shaded area in (B).}
  \label{fig:figure4}
\end{figure*}
In this experiment, we constructed hybrid coding chains in which selector and instructor roles were randomly assigned to either a human or an AI agent according to a fixed ratio (Fig. \ref{fig:figure4}A). As in all other experiments, human and AI instructors each interacted with an AI code generation system to create the SVG code. We tested three hybrid conditions: human-75\% (AI-25\%, where humans take 75\% of iterations and AI takes the rest), human-50\% (AI-50\%), and human-25\% (AI-75\%). 

As shown in Fig. \ref{fig:figure4}B, all hybrid conditions performed better than fully AI-led vibe coding ($\Delta=.55$, $p<.001$, $d=.79$, for human-75\%; $\Delta=.34$, $p=.003$,  $d=.481$, for human-50\%; $\Delta=.17$, $p=.056$, $d=.27$, for human-25\% compared with the original AI condition), showing that even limited human involvement improves performance. However, performance decreased steadily as the proportion of AI increased:
human-75\% condition (iterations 13-15:  $M=2.23$,  $\text{CI}_{95\%}$=[2.05, 2.43]; human-50\%: $M=2.02$,  $\text{CI}_{95\%}$=[1.83, 2.22],  and human-25\% conditions ($M=1.85$, $\text{CI}_{95\%}$=[1.70, 2.01]), showing a trade-off between the human proportion and the coding performance ($r=-.97$, $p=.006$; Fig. \ref{fig:figure4}C). Moreover, none of the hybrid conditions showed positive improvement with iterations (correlations for human-75\%, 50\% and 25\% were $r=-.07$, $-.12$, $-.23$, respectively).

\subsection{Role division in vibe coding}\label{results:role division}
Beyond the question of whether vibe coding generally benefits from human-AI hybrid teams, we next asked \textit{how} roles should be allocated between different agents across iterations. In our framework, we distinguish between two roles -- selectors and instructors -- reflecting common practices in real-world collaborative programming, where humans are involved both in choosing among alternatives and in providing guidance. To assess the respective contributions of these roles, we conducted an ablation study in which the selection step was removed and the edited output was always propagated to the next iteration.

We recruited 31 new human participants for 20 coding chains (experiment 10 in Appendix Table \ref{tab:exp_counts}). As shown in Fig. \ref{fig:figure5}A (light red), removing the selection step produced a significant reduction in human vibe-coding quality relative to the original human condition with selection in the final iteration (iterations 13-15:  $M=2.26$, $\text{CI}_{95\%}$=[2.09, 2.42], which is significant compared with original human condition ($\Delta=-.44$, $p<.001$, $d=-.62$).
This shows that when humans guide the process, explicit evaluation and choice between alternatives help maintain a coherent direction and support cumulative improvement.
However, for AI as shown in Fig. \ref{fig:figure5}A (cyan), removing the selection had almost no effect (iterations 13-15: $M=1.68$, $\text{CI}_{95\%}$=[1.56, 1.81], compared with the original AI condition $\Delta=.05$, $p=.339$, $d=.08$) suggesting that AI-generated guidance already follows a fixed internal preference, and explicitly comparing alternatives does not improve its trajectory. 
\begin{figure}[H]
  \begin{center}
    \includegraphics[width=0.95\columnwidth]{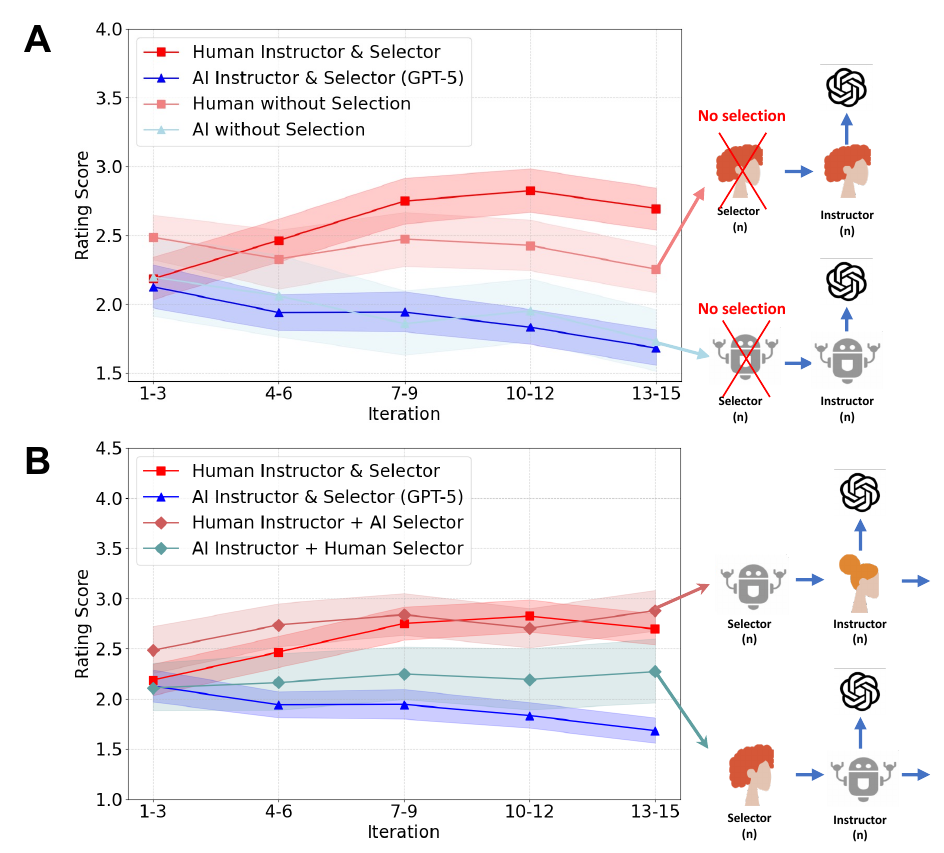}
  \end{center}
  \caption{Role division for collaborative vibe coding. (A) Validation results of experiments removing the code selection step. (B) Validation results of experiments setting human/AI agents in different roles (instructor/selector).}
  \label{fig:figure5}
\end{figure}
Fig. \ref{fig:figure5}B (light red) presents the results of a complementary ablation study with 78 participants in which the AI system assumed the selection role while humans provided the instruction (experiment 12 in Appendix Table \ref{tab:exp_counts}). Replacing the human selector with an AI selector, while retaining a human instructor, yielded performance comparable to fully human-dominated vibe coding (iterations 13-15: $M=2.88$,  $\text{CI}_{95\%}$=[2.68, 3.08], compared with the original human condition $\Delta=.18$, $p=.081$, $d=.23$) suggesting that selection can be offloaded to AI with little loss in performance, offering a practical way to reduce human effort without sacrificing the quality of collaborative vibe coding.

This further suggests that human instructors tend to generate effective improvements. Moreover, human selection becomes essential when the instructor is an AI system (experiment 13 in Appendix Table \ref{tab:exp_counts}, $M=2.27$, $\text{CI}_{95\%}$=[1.96, 2.59], compared with the original AI condition $\Delta=.59$, $p<.001$, $d=.77$). Taken together, these findings highlight the importance of human supervision in mitigating compromised or suboptimal AI-generated outputs. Further analysis of the selection acceptance rate (i.e., whether the selector chooses the current SVG) supports this by revealing misalignment between humans and AI when collaborating on a task. Specifically, the acceptance rate when humans act as both instructor and selector ($M=.66$, $\text{CI}_{95\%}=[.62, .70]$) is significantly higher than the condition with a human instructor and an AI selector ($M=.38$, $\text{CI}_{95\%}=[.33, .42]$, difference $\Delta=.28$, $p<.001$, $d=.59$) and vice versa ($M=.34$, $\text{CI}_{95\%}=[.30, .39]$, $\Delta=.32$, $p<.001$, $d=.67$).

\subsection{Robustness to AI model type and social information}\label{results:robustness} 
\begin{figure}[H]
  \begin{center}
    \includegraphics[width=0.8\columnwidth]{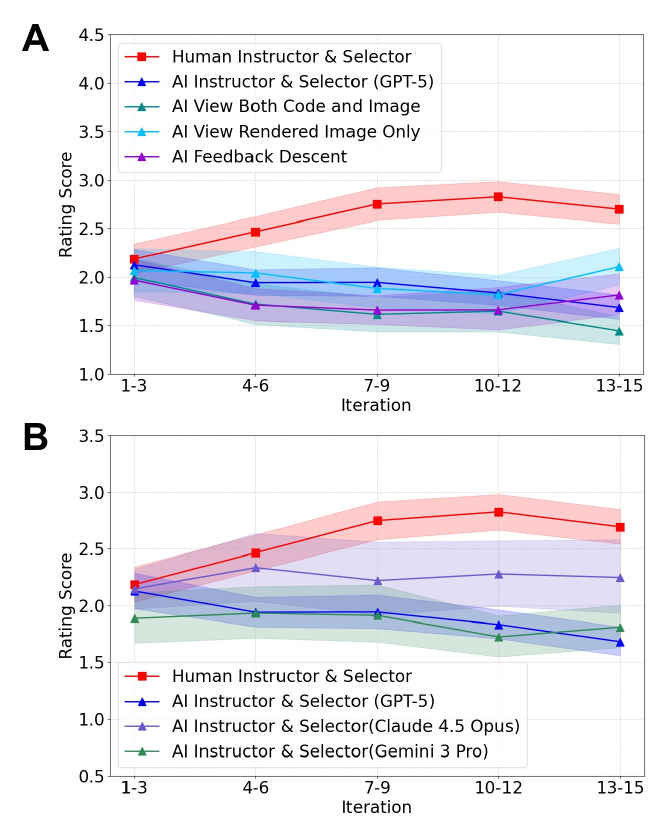}
  \end{center}
  \caption{Control experiments for vibe coding pipeline. (A) Experiment results of different viewing modes in AI-led vibe coding, with GPT-5 acting as instructors and selectors. (B) Results of AI-led vibe coding, under different AI systems.}
  \label{fig:controlsAI}
\end{figure}

To test the robustness of our experimental vibe coding pipeline to other factors, we conducted several control experiments. First, we changed how the AI vibe coder evaluated the SVG outputs (Fig. \ref{fig:controlsAI}A). In the main experiment, the AI vibe coder observed the SVG code directly. In the control conditions, we instead allowed the AI vibe coder to view the rendered SVG image only, matching the information available to human vibe coders, or to view both the SVG code and its rendered image. Across both viewing modes, AI-led vibe coding showed similar performance to the main AI condition, remaining significantly worse than human-led vibe coding ($\Delta=-.59$, $-1.25$, both $p<.001$) and exhibiting no improvement across iterations (Only SVG-rendered image visible: $r=-.01$, $\text{CI}_{95\%}$=[-.17, .15]; Both SVG code and rendered image visible: $r=-.32$, $\text{CI}_{95\%}$=[-.46, -.18]). This indicates that the AI’s failure is not due to limited access to visual and code information.

We also tested an advanced recent AI pipeline, named FeedBack Descent \citep{lee2025feedback}, where the selector not only provided the selection but also gave feedback on the code (for more details, see Appendix \ref{appendix:ai-vibe-coding}). Similarly, there was no improvement across iterations ($r=-.08$, $\text{CI}_{95\%}$=[-.26, .09]) and the final performance is significantly lower than the human-led condition ($\Delta=-.88$, $p<.001$, $d=-1.32$; Fig. \ref{fig:controlsAI}A).
We further tested whether the observed effects depend on the specific AI model used for vibe coding (Fig. \ref{fig:controlsAI}B). In addition to GPT-5, we repeated the AI-led experiments using Claude-4.5-Opus (Anthropic, \cite{anthropic2025opus}) and Gemini-3-Pro (Google, \cite{team2023gemini}) as instructor and selector, while keeping the code generator to be GPT-5 (to facilitate a fair comparison between the AI model performance). 
Both alternative models produced lower final similarity scores than human-led vibe coding (Claude: $M=2.25$, $\text{CI}_{95\%}$=[1.93, 2.59]; Gemini: $M=1.81$, $\text{CI}_{95\%}$=[1.63, 2.01]) and showed a similar pattern of no improvement across rounds (Claude: $r=.02$, $\text{CI}_{95\%}$=[-.14, .18]; Gemini: $r=-.08$, $\text{CI}_{95\%}$=[-.23, .07]; Fig. \ref{fig:controlsAI}B). These results demonstrate that the performance gap and collapse effect are not specific to a single model, but instead reflect a general limitation of current AI systems when acting as high-level guidance providers in multi-round collaborative coding. As the performance gap was getting somewhat smaller with a more advanced AI model (Claude vs. human-led: $\Delta=-.45$, $p=.005$, $d=-.56$), our results also indicate that the improvement of AI may enhance the alignment between AI and human behavior.

\section{Discussion}


As vibe coding becomes an increasingly widespread practice, it is important to characterize how it impacts collective outcomes, and what arrangements of human-AI collaboration yield the best outcome. Our experimental paradigm makes it possible to systematically explore different human-AI compositions and roles in vibe coding. Across 18 experiments (and two evaluation experiments), human-led coding consistently improved over iterations, while AI-led coding often collapsed despite access to the same information and similar execution capabilities. This highlights significant challenges for the AI systems we studied in maintaining coherent high-level cumulative improvements across repeated interactions, a critical factor for successful vibe coding. Our results show how studying human–AI collaboration directly reveals behaviors that would be missed by evaluating autonomous AI agents alone.

One of our key findings is the systematic misalignment between how humans and AI systems use language to guide collaboration. Humans tend to provide short, action-oriented instructions that focus on what requires changing, while AI systems produce long, descriptive instructions that emphasize details. Importantly, constraining instruction length did not eliminate this gap, indicating that misalignment arises from differences in strategy rather than verbosity. This is also supported by our results that AI systems have lower sensitivity to the quality difference between human- and AI-led vibe coding creations, alongside a tendency to evaluate their own creations more favorably. This aligns with broader concerns that large language models may optimize for descriptive completeness instead of goal-directed coordination in interactive tasks \citep{tsvilodub2023overinformative,miehling2024language,ma2025pragmatics}.

Our hybrid and role-division experiments further clarify how labor could be effectively allocated in human–AI systems. Partial human involvement improved performance relative to fully AI-driven pipelines, but even limited AI guidance could shift the coding trajectory in ways that later human input could not fully correct. At the same time, decomposing the vibe coder role showed that high-level idea generation and instruction are -- in our setup -- the critical human contributions, whereas evaluation and selection can often be delegated to AI without loss in performance. This suggests a practical design principle for hybrid systems: humans should set direction, while AI can support evaluation and execution.

Advances in contemporary AI systems are proceeding at an extraordinary pace. In principle, coding agents specifically optimized for tasks such as SVG image generation can achieve high levels of performance. Indeed, in our experiments, the model we evaluated matched human participants at the outset (iteration 1). Much of current engineering efforts are directed toward maximizing performance at this initial stage. However, when performance is examined at the level of group dynamics across successive iterations, a different pattern emerges: AI-led systems tend to deteriorate over time rather than improve. Although further task-specific optimization -- such as models tailored explicitly for SVG generation -- may substantially reduce or even eliminate the overall performance gap relative to humans, important insight of our work concerns the critical contrast between strong initial performance and subsequent degradation. As vibe coding becomes more widespread and is deployed for increasingly complex tasks, understanding and addressing this divergence will be essential, highlighting the broader implications of our findings.

More broadly, our findings contribute to a cognitive science of hybrid societies by showing that collective performance depends not only on agent competence, but on how roles, feedback, and control are assigned and structured over time between humans and machines~\citep{bhatt2025should}. Vibe coding offers a tractable model system for studying these dynamics, with implications beyond programming, toward other creative design work and decision-making in human–AI teams \citep{amershi2019guidelines,bansal2021does,rezwana2023designing,davis2025co, collins2024building}.

\subsection{Limitations}
Several limitations of our study should be noted. First, in our task, participants provided improvement instructions only after selecting the preferred SVG, whereas in real-world coding contexts, developers typically also articulate the shortcomings of the rejected alternative. Initial tests implementing a recent feedback-based method from the literature \citep{lee2025feedback} did not yield substantial performance gains (Figure~\ref{fig:controlsAI}). Nevertheless, future work should more systematically explore alternative feedback paradigms; for example, structured code-review procedures modeled on open-source development practices may provide richer supervisory signals. Second, although SVG coding is intuitive and accessible, it remains a simplified task, and future studies should test more realistic coding settings, such as website or game development. Third, we used a single AI model as either a coding assistant, instructor, or selector, while our prior work has explored mixtures of different AI systems~\citep{shiiku2025dynamics}; whether combining multiple AI agents improves vibe coding remains an open question. Finally, our experiments model collaboration as a linear iterative chain, whereas real-world coding often involves team-based, parallel collaboration, which should be explored in future work \citep{shiiku2025dynamics, marjieh2025characterizing}. We also note that we only study short-term human-AI interaction effects, but future work should also consider long-term feedback loops and risks like cognitive decline and overreliance~\citep{oktar2025identifying,ibrahim2025measuring,collins2025revisiting}. 

\subsection{Conclusion}

Understanding the potential impacts of AI systems on human society is going to require empirical work that explores the behavior of human-AI hybrid systems. In this work, we introduced a controlled experimental framework for studying collaborative vibe coding and used it to systematically compare human- and AI-led processes. We show that humans provide uniquely effective high-level guidance across iterations and that AI guidance often leads to performance collapse. Additionally, we find that a role allocation that keeps humans in charge of direction while offloading evaluation to AI can improve hybrid performance. Together, these results demonstrate the value of studying human–AI collaboration directly and offer concrete insights for designing effective hybrid human–AI systems.

\section{Acknowledgments}
This work was supported by the NSF grant ``Collaborative Research: Research Infrastructure: HNDS-I: Building Infrastructure to Study Human-AI Hybrid Societies in Experimental Social Networks'' (Award BCS-2523500) and partially supported by the NSF grant ``Collaborative Research: Designing smart environments to augment collective learning \& creativity'' (Award BCS-2421386). KMC acknowledges support from the NSF SBE SPRF. ChatGPT version 5.2 (OpenAI) was used to assist manuscript editing and proofreading. Authors reviewed each of the edit suggestions, and approved the final version.

\backmatter

\bibliography{cogsci_bibliography_template}
\newpage
\begin{appendices}
\startcontents[appendices]
\printcontents[appendices]{}{0}{\section*{Appendix Contents}}

\section{Supplementary Methods}\label{smethods}
\subsection{Participants and AI Queries}

\begin{table}
\caption{Experiment Conditions: Instructor and Selector Roles}\label{tab:exp_roles}
\footnotesize
\begin{tabular}{@{}lcc@{}}
    \toprule
    \textbf{Experiment Condition} & \textbf{Instructor} & \textbf{Selector} \\
    \midrule
    1. Human-Led & Human & Human \\
    2. AI-Led & AI (GPT-5) & AI (GPT-5) \\
    3. Human-Led (Asocial) & Human & Human \\
    4. AI-Led: Instruction Limit 10 Words & AI (GPT-5) & AI (GPT-5) \\
    5. AI-Led: Instruction Limit 20 Words & AI (GPT-5) & AI (GPT-5) \\
    6. AI-Led: Instruction Limit 30 Words & AI (GPT-5) & AI (GPT-5) \\
    7. Hybrid: Human 75\% - AI 25\% & Hybrid & Hybrid \\
    8. Hybrid: Human 50\% - AI 50\% & Hybrid & Hybrid \\
    9. Hybrid: Human 25\% - AI 75\% & Hybrid & Hybrid \\
    10. Human-Led: No Selection & Human & - \\
    11. AI-Led: No Selection & AI (GPT-5) & - \\
    12. Human-Instructor+AI-Selector & Human & AI (GPT-5) \\
    13. Human-Selector+AI Instructor & AI (GPT-5) & Human \\
    14. AI-Led: View Rendered Image Only & AI (GPT-5) & AI (GPT-5) \\
    15. AI-Led: View Both Code and Image & AI (GPT-5) & AI (GPT-5) \\
    16. AI-Led: Feedback Descent & AI (GPT-5) & AI (GPT-5) \\
    17. AI-Led: Gemini-3-Pro & AI (Gemini-3-Pro) & AI (Gemini-3-Pro) \\
    18. AI-Led: Claude-4.6-Opus & AI (Claude-4.5-Opus) & AI (Claude-4.5-Opus) \\
    19. Evaluation Rating (Human) & - & - \\
    20. Evaluation Rating (AI/GPT-5) & - & - \\
    \botrule
\end{tabular}
\end{table}

\begin{table}
\caption{Experiment Information}\label{tab:exp_counts}
\footnotesize
\begin{tabular}{@{}lcccccccc@{}}
    \toprule
    \textbf{Exp.}\footnotemark[1] & \textbf{N\textsubscript{h}}\footnotemark[2] & \textbf{N\textsubscript{b}}\footnotemark[3] & \textbf{N\textsubscript{t/h}}\footnotemark[4] & \textbf{N\textsubscript{ch}}\footnotemark[5] & \textbf{N\textsubscript{q}}\footnotemark[6] & \textbf{N\textsubscript{tr}} (exp/inv)\footnotemark[7] & \textbf{Age}\footnotemark[8] & \textbf{Sex}\footnotemark[9] \\
    \midrule
    Exp. 1  & 49  & -   & 10  & 30 & 450   & 450/2   & 42.40$\pm$12.10 & 20/27/2  \\
    Exp. 2  & -   & 450 & -   & 30 & 1290  & 450/0   & -               & -        \\
    Exp. 3  & 10  & -   & 45  & 30 & 450   & 450/3   & 36.80$\pm$11.18 & 3/7/0    \\
    Exp. 4  & -   & 450 & -   & 30 & 1290  & 450/0   & -               & -        \\
    Exp. 5  & -   & 450 & -   & 30 & 1290  & 450/0   & -               & -        \\
    Exp. 6  & -   & 450 & -   & 30 & 1290  & 450/0   & -               & -        \\
    Exp. 7  & 221 & 76  & 1   & 20 & 440   & 300/4   & 47.62$\pm$18.05 & 90/129/2 \\
    Exp. 8  & 143 & 153 & 1   & 20 & 595   & 300/6   & 51.03$\pm$20.40 & 62/81/0  \\
    Exp. 9  & 85  & 212 & 1   & 20 & 692   & 300/8   & 54.10$\pm$22.71 & 34/51/0  \\
    Exp. 10 & 31  & -   & 10  & 20 & 300   & 300/1   & 43.43$\pm$13.15 & 15/15/0  \\
    Exp. 11 & -   & 150 & -   & 10 & 300   & 150/0   & -               & -        \\
    Exp. 12 & 46  & 260 & 10  & 20 & 560   & 300/1   & 43.98$\pm$13.69 & 22/24/0  \\
    Exp. 13 & 32  & 147 & 5   & 10 & 297   & 150/4   & 43.84$\pm$13.09 & 15/17/0  \\
    Exp. 14 & -   & 150 & -   & 10 & 430   & 150/2   & -               & -        \\
    Exp. 15 & -   & 150 & -   & 10 & 430   & 150/2   & -               & -        \\
    Exp. 16 & -   & 150 & -   & 10 & 430   & 150/4   & -               & -        \\
    Exp. 17 & -   & 150 & -   & 10 & 430   & 150/0   & -               & -        \\
    Exp. 18 & -   & 150 & -   & 10 & 430   & 150/0   & -               & -        \\
    Exp. 19 & 120 & -   & 175 & -  & -     & 21000/0 & 40.93$\pm$12.69 & 51/68/1  \\
    Exp. 20 & -   & 2700 &     & -  & 2700   & 2700/0   & -               & -        \\
    \midrule
    \textbf{Summary} & \textbf{737} & \textbf{6248} & & \textbf{350} & \textbf{14094} & \textbf{28950/39} & & \\
    \multicolumn{3}{@{}l}{\textbf{Invalid Rate (\%)\footnotemark[10]}} & & & & \textbf{.135} & & \\
    \multicolumn{3}{@{}l}{\textbf{Invalid Rate Main (\%)\footnotemark[11]}} & & & & \textbf{.743} & & \\
    \botrule
\end{tabular}
\footnotetext[1]{The number represents the experiment number in Appendix Tab.~\ref{tab:exp_roles}.}
\footnotetext[2]{The number of unique human participants.}
\footnotetext[3]{The number of iterations an AI agent performed as a participant (where one iteration corresponds to a single role as instructor, or as both instructor and selector). In Exp. 12 \& 13 this represents the number of bot selector or instructor.}
\footnotetext[4]{The number of trials expected to be completed per participant.}
\footnotetext[5]{The number of chains in the experiment.}
\footnotetext[6]{The total number of API queries (instructor + selector + code generation).}
\footnotetext[7]{exp: total experimental trials; inv: number of invalid/excluded trials.}
\footnotetext[8]{Age is reported as mean $\pm$ SD; dash indicates AI-only condition with no human participants.}
\footnotetext[9]{Self-reported sex is reported as male/female/other; dash indicates AI-only condition.}
\footnotetext[10]{Invalid rate across all experiments (Exp. 1--20).}
\footnotetext[11]{Invalid rate for main experiments (Exp. 1--18) only.}
\end{table}

In this part we present the information of participants as well as AI involved in each experiment. All details can be found in Table \ref{tab:exp_roles} and \ref{tab:exp_counts}.

All participants provided consent in accordance with the Cornell University–approved protocol IRB0148995. All participants were recruited at online recuiting platform Prolific (\href{https://www.prolific.com/}{https://www.prolific.com/}) and were pre-selected by the following properties: 1) born and resident in the United Kingdom; 2) the first language is English; 3) using a laptop with Google Chrome Browser installed. Participants were compensated with a rate of \pounds 10 per hour according to the progress of their experiment. A total of 737 participants were recruited for 9 human-involved experiments after excluding those who didn't provide consent or failed to provide any valid data.


For AI queries, all api requests were made through an integrative platform OpenRouter (\href{https://openrouter.ai/}{https://openrouter.ai/}). A total of 14,094 AI queries were made for different tasks of 20 experiments following the instruction in Appendix. \ref{appendix:prompts}. Three closed-source models were involved in our experiment: 1) GPT-5 from OpenAI \citep{singh2025openai}; 2) Gemini-3-Pro from Google \citep{team2023gemini} \citep{anthropic2025opus} and 3) Claude-4.5-Opus from Anthropic \citep{anthropic2025opus}. These three models are considered as the most representative and advanced AI models at the stage of this study, ensuring the stability of our discoveries.

\subsection{Experiment Implementation}\label{appendix:exp-implement}

\subsubsection{Data Availability}
All experimental materials, raw data, and analysis scripts are publicly available in the OSF repository associated with this project \citep{hu_marjieh_collins_sucholutsky_jacoby_2026}.

\subsubsection{PsyNet}
All experiments were implemented using PsyNet (\href{http://www.psynet.dev}{http://http://www.psynet.dev}), an online framework for behavioral experiments \citep{harrison2020gibbs}. PsyNet supports interactions between humans and AI agents within social networks \citep{shiiku2025dynamics,marjieh2025characterizing}.

\subsubsection{Human-Involved Vibe Coding}\label{appendix:human-vibe-coding}
In this section, we describe in detail the procedure for each human-involved experiment, as documented in Table \ref{tab:exp_roles} and \ref{tab:exp_counts}. These tables also summarize the experimental design, number of human participants, number of AI calls, and participants’ demographic characteristics.

\textbf{Human-Led Experiment} (Exp. 1, Results \ref{results:human-ai main comparison}). Under this condition, we recruited human participants as both instructors and selectors. 10 unique animal images (cat, dog, tiger, bird, elephant, penguin, shark, zebra, giraffe, and panda) were utilized in the study. In each experiment, there would be one image selected as the reference, and there would be 15 iterations per chain (same for most experiments in this study unless specified). Each image was repeated three times, resulting in a total of 30 independent chain experiments. In the initial phase of the first two iterations, when fewer prior images were available compared to later stages, we recruited only instructors to perform the initialization and the first round of SVG code editing. Each iteration only involved one participant, and a participant was expected to finish 10 iterations of different chains. At the end of the experiment, there were a few survey questions collecting participants' demographic information, including their age and gender, their strategy in the experiment, and any technical problems they encountered. Theoretically, 45 participants could fulfill the requirement of this experiment and produce 450 SVG codes. However, due to the early quitting of some participants, we actually recruited 49 participants, creating 448 valid SVG codes with 2 missing codes due to failed trials or, in rare cases, the generation of syntactically invalid SVGs by the code generator. For the convenience of evaluation image allocation, missing SVG codes in our study were replaced by randomly sampled SVG codes with a special mark (e.g., set the iteration number as 16) and excluded from our final rating analysis. The proportion of syntactically invalid SVGs produced by the code generator was low (below 1\%) across all experiments and is reported for each experiment in Table \ref{tab:exp_counts}.

\textbf{Human-Led Experiment (Asocial)} (Exp. 3, Results \ref{results:human-ai main comparison}). In this experiment, each participant independently completed all 10 iterations of a single chain before moving on to the next chain. The procedure therefore resembles repeated interaction between a single individual and a coding agent (i.e., an asocial setting), rather than the socially distributed, multi-participant chains used in most of our experiments.In total, there were 30 chains, with each animal image repeated three times. Each participant completed three chains, each corresponding to a different animal reference.

\textbf{Hybrid Experiment} (Exp. 7-9, Results \ref{results:hybrid}). In the hybrid setting, an approximate proportion of iterations were replaced by AI instructors and selectors randomly in chains according to the rate we set in advance. We alternated between recruiting human participants and AI agents to maintain the target proportion as closely as possible (Table \ref{tab:exp_counts}). In our study, we explored three hybrid settings: human-75\% (AI-25\%), human-50\% (AI-50\%) and human-25\% (AI-75\%). In each experiment there were 20 chains with each animal repeating twice and each participant was expected to complete one experimental trial followed by survey questions as above. 
Note that in the main experiment participants initially took part in ten trials. However, we found empirically that performance did not change across these repeated trials. To facilitate balanced recruitment of both human participants and AI bots, we therefore limited each human participant and each AI bot to participating only once.

\textbf{Human-Led Experiment - No Selection} (Exp. 10, Results \ref{results:role division}). We removed the selection step for this experiment and kept every other steps consistent with the main human-led experiment. In this case, images from the current generation are always propagated to the next generation. 20 chains were deployed and each participant was expected to complete 10 trials across chains and the survey. 

\textbf{Human as Instructor \& AI as Selector} (Exp. 12, Results \ref{results:role division}). In this experiment, human participants served as instructors. Selection was performed by the AI immediately after participants submitted their instructions. There were 30 chains in the experiments and each participant was expected to complete 10 trials across chains, in addition to the survey.

\textbf{Human as Selector \& AI as Instructor} (Exp. 13, Results \ref{results:role division}). As AI acted as the instructor in this condition, we ran the first two trials locally to ensure the start of the experiment would not get stuck due to the delay of API queries. Therefore, only 13 trials were collected for each chain and 10 chains were deployed.  

\subsubsection{AI-Led Vibe Coding}\label{appendix:ai-vibe-coding}
\textbf{AI-led or AI-involved experiments} (Exp. 2, 4-6, 7-9, 11, 17-18, Results \ref{results}). In all AI-led or AI-involved experiments , each AI trial was finished by an individual API query to OpenRouter (\href{https://openrouter.ai/}{https://openrouter.ai/}). Here, we rely on the native integration of AI agents within the PsyNet experimental framework to execute model calls, enabling us to faithfully replicate the full procedure of the human experiments. AI was asked to act as different roles (instructor or selector) by setting the prompt in Appendix \ref{appendix:prompts}. AI instructors or selectors were provided with the reference image as PNG file. To imitate a naturalistic coding procedure of AI agents, selections and instructions were based on the unrendered SVG codes directly. To test the influence of observing images directly we performed controlled experiment  (Exp. 14 and Exp. 15, see below). Most experiments were conducted using GPT-5 between November 2025 and February 2026. This model version does not expose the temperature parameter, so all runs were performed using the default settings. An exception is Fig. \ref{fig:controlsAI}, where we used different models as instructors. Specifically, we used Claude 4.5 Opus and Gemini 3 Pro accessed through API calls with default settings. To facilitate comparison across conditions, the code-generation system was kept constant and implemented using GPT.

\textbf{AI-Led Experiments: Changing Available Information for AI} (Exp. 14-15, Results \ref{results:robustness}). In Exp. 14, AI instructors and selectors were provided only with rendered images and compared those with the reference image. This was designed to match human-led experiment where human could only check the output of the coding process and make suggestion to improve. And in Exp. 15, AI instructors were presented with both SVG codes and rendered images which fully considered the advantage of AI agents. The results provided in Figure \ref{fig:controlsAI} confirm that these manipulations did not produce a significant effect, indicating that the specific modality presented to the AI was unlikely to account for the observed reduction in performance.

\textbf{AI-Led Experiment: Feedback Descent} (Exp. 16, Results \ref{results:robustness}). In the AI-led experiment implementing Feedback Descent (Exp. 16), we adopted the procedure described by \citep{lee2025feedback}, a recent framework for feedback-driven code optimization. The overall design parallels that of our other AI-based experiments, with two modifications relative to experiment (Exp. 2). First, a crtieria for evaluations called ``rubric'' was added to make sure the selector to evaluate the generated codes following a certain standards. The rubric was designed for the purpose of helping the selector to evaluate: 1) How well an SVG illustration matched the original image; 2) What specific aspects to look for (proportions, colors, shapes, details, etc.) and 3) What made a good match vs. a poor match. 
Here we generated the rubric specifically for each animal ahead of the experiment by GPT-5.  Appendix~\ref{appendix:prompt-rubric} describes the procedure used to generate the image-specific rubric (which itself implement as a call to GPT). In addition, Appendix~\ref{appendix:example-rubric} provides two illustrative examples. In this experiment,  the selector was requested to provide feedback other than the selection pointing out why the selector was better and how to improve the code as specified in Appendix \ref{appendix:prompt-ai-instructor-selector}. The feedback was then provided to the code generator together with the instruction from AI instructor to modify the SVG code. The results shown in Figure~\ref{fig:controlsAI} confirm that these procedural differences did not yield substantial improvement, while adding further complexity to the experimental setup.

\subsubsection{Evaluation Experiments\label{appendix:evaluation}}

\textbf{Post-Experiment SVG Processing}. All experimental SVG codes were rendered into Portable Network Graphics (PNG) images through a python package CairoSVG (\href{https://cairosvg.org/}{https://cairosvg.org/}). A total of 5,211 SVG codes were successfully rendered into PNG images. A small fraction of images (.743\%) produced SVG files that were incompatible with CairoSVG, likely due to the code generator occasionally emitting syntactically invalid SVG. To maintain a balanced evaluation set, we replaced these 39 SVGs (out of 5,250 rendered images) with randomly sampled images from the successfully rendered pool. The substituted cases were explicitly flagged and excluded from subsequent analyses. All processed PNG files can be found in  \citep{hu_marjieh_collins_sucholutsky_jacoby_2026}. 


\textbf{Human Evaluation} (Exp. 19, Results \ref{results}). According to Table \ref{tab:exp_counts}, a total of 350 chains were deployed for the whole study with each animal category assigned to 35 chains. To facilitate comparison, each participant was assigned to a single animal category for the entire experiment. For evaluation, 5 images were randomly sampled from each chain, yielding 175 trials per participant. The order of chains presented to evaluators was randomized, and within each chain, the 5 trials from different iterations were also presented in randomized order. In total, 120 participants were recruited, contributing 21,000 valid rating scores.

\textbf{AI Evaluation} (Exp. 20, Results \ref{results:human-ai main comparison}).In the AI evaluation experiment, each assessment was performed via an individual API query, with the AI processing all rendered images in randomized order. To enhance result stability, the entire evaluation procedure was repeated 3 times, yielding a total of 2,700 AI-generated rating scores.

\subsection{Rating Data Analysis}\label{appendix:rating-data-analysis}

Because image selection during validation involved random sampling, each image received approximately 3 ratings (with at least 1 rating guaranteed per image). Although obtaining a stable estimate for an individual image would ideally require 5–10 ratings, our analyses were conducted at the aggregate level—collapsing across images—rendering this sampling density sufficient.

As the number of ratings varied slightly across images and our analysis focused on averages across many images, we computed the mean rating per image and used these averaged values in subsequent analyses.

For most analyses, and to increase statistical power, we grouped consecutive iterations into bins of three within each experimental condition and image category (iterations 1–3, 4–6, 7–9, 10–12, 13–15), excluding missing values. This procedure partitioned all rating data into five sequential bins for analysis.

\subsubsection{Bootstrapping procedure}
We compared mean values (e.g., ratings from human evaluations) using a bootstrap procedure. We generated 10,000 bootstrap samples by sampling with replacement from the data and computed the mean for each sample. Confidence intervals were obtained from the 2.5th and 97.5th percentiles of the resulting bootstrap distribution. The same procedure was used to compute the confidence intervals shown in the main figures, where means were calculated over binned iterations (e.g., iterations 1–3).

\subsubsection{Between-Condition Analysis}\label{appendix:bootstrap}

We used similar procedure to compared performance between conditions (e.g., human vs. AI). To account for possible differences in the number of observations, following \cite{davison1997bootstrap}, we first concatenated the data from the two conditions and then repeatedly sampled from the combined dataset with replacement. For each resampled dataset, we computed the mean difference between conditions across 10,000 bootstrap iterations and compared these values to the observed difference. Namely, we first computed the observed data:
\begin{equation}
\Delta_{\text{obs}} = \bar{X}^{\text{(Cond2)}} - \bar{X}^{\text{(Cond1)}}
\end{equation}
where $\bar{X}^{\text{(Cond1)}}$ and $\bar{X}^{\text{(Cond2)}}$ are the mean of all pooled values of the first and second conditions, respectively. 

For each bootstrap iteration $b$ (where $b = 1, 2, \ldots, 10{,}000$),
we then computed the bootstrapped difference: 
\begin{equation}
    \Delta^{(b)} = \bar{X}^{(\text{Cond2}, b)} - \bar{X}^{(\text{Cond1}, b)}
\end{equation}
where $\bar{X}^{(\text{Cond2}, b)}$ and $\bar{X}^{(\text{Cond1}, b)}$ were calculated using sampled data from the combined dataset.

One-sided p-values were obtained by comparing the observed difference to the bootstrap distribution and calculating the proportion of bootstrap samples that were greater than (or smaller than) the observed value.

\begin{equation}
p = \frac{1}{10{,}000} \sum_{b=1}^{10{,}000} \mathbbm{1}(\Delta^{(b)} \leq \Delta_{\text{obs}})
\end{equation}

 Effect size was quantified using Cohen's $d$ for independent samples:
\begin{equation}
d = \frac{\bar{X}^{\text{(Cond2)}} - \bar{X}^{\text{(Cond1)}}}{s_{\text{pooled}}}
\end{equation}
where the pooled standard deviation is:
\begin{equation}
s_{\text{pooled}} = \sqrt{\frac{(n^{\text{(Cond2)}} - 1)s_{\text{Cond2}}^2 + (n^{\text{(Cond1)}} - 1)s_{\text{Cond1}}^2}{n^{\text{(Cond2)}} + n^{\text{(Cond1)}} - 2}}
\end{equation}
where $n^{\text{(Cond2)}}$ and $n^{\text{(Cond1)}}$ are the number of individual values in step pools from two experiment conditions, and $s_{\text{Cond2}}$ and $s_{\text{Cond1}}$ are the sample standard deviations.

The same procedure was applied both when using the full dataset and when restricting the analysis to specific iterations, for example, when comparing only the first three iterations or the last three iterations.

\subsubsection{Correlation Analysis}

We assessed the relationship between iteration number and rating values using Pearson correlation with bootstrap confidence intervals. We used the full iteration data (all 15 iterations) rather than step-averaged values.

\textbf{Pearson Correlation Coefficient}. For each term within a condition, we calculated:
\begin{equation}
r = \frac{\sum_{i=1}^{n}(x_i - \bar{x})(y_i - \bar{y})}{\sqrt{\sum_{i=1}^{n}(x_i - \bar{x})^2}\sqrt{\sum_{i=1}^{n}(y_i - \bar{y})^2}}
\end{equation}
where $x_i$ represents iteration numbers (1, 2, \ldots, 15), $y_i$ represents the corresponding rating values, $n$ is the number of valid data points, and $\bar{x}$ and $\bar{y}$ are the respective means.

\textbf{Statistical Significance}. The $p$-value for each correlation was computed using a $t$-test:
\begin{equation}
t = r\sqrt{\frac{n-2}{1-r^2}}
\end{equation}
which follows a $t$-distribution with $n-2$ degrees of freedom under the null hypothesis of no correlation \citep{fisher1915frequency}.

\textbf{95\% CI}. For each bootstrap iteration $b$, we firstly resample pairs $(x_i, y_i)$ with replacement, maintaining the pairing and then calculate the Pearson correlation $r^{(b)}$ for the bootstrap sample. With that the 95\% CI was determined using the percentile method:
\begin{equation}
    \text{CI}_{95\%} = [\text{P}_{2.5}(r^{(b)}), \text{P}_{97.5}(r^{(b)})]
\end{equation}

\subsubsection{Human-AI Misalignment Analysis}\label{appendix:misalignment-analysis}
We performed separate analyses for comparing 1) the misalignment between human and AI evaluation rating results in Results \ref{results:human-ai main comparison}; 2) the misalignment between human's and AI's semantic instructions within and between animal image categories in Results \ref{results:semantic}; 3) the misalignment between acceptance rate in the selection when human and AI acted as the selector in Results \ref{results:role division}.

\textbf{Human-AI Evaluation Misalignment}. In Results \ref{results:human-ai main comparison}, we provided a comparison between the rating results from human evaluators (Fig. \ref{fig:figure2-extra}) and AI evaluators (Appendix Fig. \ref{fig:ai-rating}). We performed bootstrapping (as described in Appendix \ref{appendix:bootstrap}) to compare: 1) the overall improvement in rating scores and 2) the increase of rating scores on AI-led experiments vs. human-led experiments, when evaluators were changed from humans to AI. 


\textbf{Human-AI Semantic Misalignment}. In Results \ref{results:semantic}, we compared the misalignment between human- and AI-led vibe coding at the semantic level. As described in Results \ref{results:semantic}, we first converted all instructions from human- and AI-led experiments (Exp. 1, 2 and 4-6 in Table \ref{tab:exp_roles}) into sentence embeddings and then compared two aspects: within-category similarity and across-category similarity.

For within-category comparison, we calculated the pair-wise cosine similarities between different embeddings within the same animal category and concatenated those 10 categories' results into similarity sequences. With sequences from different experiment conditions, we utilized the same bootstrap procedure as Appendix \ref{appendix:bootstrap} for the similarity differences between human-led vibe coding and other AI-led experiments.

For across-category comparison, we calculated an averaged embedding for each category and computed cosine similarities between different category-level embeddings to get a similarity sequence, on top of which we assessed the differences in across-category similarities between experiment conditions using the same bootstrap as above.

\textbf{Human-AI Selection Misalignment}. In Results \ref{results:role division}, we evaluated the difference of acceptance rate when humans led the vibe coding or humans collaborated with AI agents. Here we computed the overall acceptance rate via:
\begin{equation}
    \text{Acceptance Rate}=\frac{\sum_{n=3}^{N} \mathbbm{1}[\text{Selection}=\text{SVG}_{n-1}]}{N-2}
    \label{eq:accept-rate}
\end{equation}
where $N=15$ was the iteration number of chain experiments, $\mathbbm{1}[\text{Selection}=\text{SVG}_{n-1}]$ is an indicator checking if the selector chooses the SVG generated from the last iteration (counted as an accepted selection). The starting was set as $n=3$ because selection appeared from the third iteration. To compare the differences in acceptance rates between human-led and collaborative vibe coding, we convert the accepted/denied selection into sequences for each condition (accepted selection was mapped to 1 and denied selection was mapped to 0). On top of that, we performed the bootstrap as Appendix \ref{appendix:bootstrap} on sequences.

\subsection{Semantic Analysis}\label{appendix:semantic-analysis}
In Results \ref{results:semantic}, we compared detailed semantic properties of human- and AI-produced instructions, involving experiments with AI-instruction limitations. We collected a total of $N_I=2646$ valid instructions from Exp. 1, 2 and 4-6 in Table \ref{tab:exp_roles}. Here is how we conduct the analysis for different metrics:

\textbf{Type-Token Ratio (TTR)}~\citep{mccarthy2010mtld} indicates lexical diversity as the ratio of unique word types to the total number of tokens in an instruction:
\begin{equation}
    \text{TTR} = \frac{|\mathcal{V}_d|}{|d|}
\end{equation}
where $|\mathcal{V}_d|$ is the number of unique word types and $|d|$ is the total token count in instructions $d$. Higher TTR reflects greater lexical variety.

\textbf{Content Ratio}~\citep{manning2008introduction} quantifies the proportion of content-bearing words (i.e., non-stopwords) relative to all tokens:
\begin{equation}
    \text{Content Ratio} = \frac{|d_{\text{content}}|}{|d|}
\end{equation}
where $|d_{\text{content}}|$ shows the number of tokens remaining after stopword (in our analysis we used stopword set from scikit-learn (\href{https://scikit-learn.org/stable/}{https://scikit-learn.org/stable/})) removal. A higher value indicates more informationally dense language.

\textbf{Topic Entropy}~\citep{deerwester1990indexing} captures the diversity of topics in the instruction. We first decomposed the embedding matrix $\mathbf{X} \in \mathbb{R}^{N \times D}$ (here $D=1536$ was the dimension of a sentence embedding encoded from the instruction) via truncated Singular Value Decomposition (SVD) into a low-rank topic representation $\mathbf{Z} \in \mathbb{R}^{N \times K}$, where $K=10$ was the number of latent topics we set in our analysis according to the animal categories we had. For each document $i$, the topic distribution was estimated from the squared loadings:
\begin{equation}
    p_k^{(i)} = \frac{z_{ik}^2}{\sum_{k'} z_{ik'}^2}, \quad
    H^{(i)} = -\sum_{k=1}^{K} p_k^{(i)} \log_2 p_k^{(i)}
\end{equation}
and normalized to $[0,1]$ by $\tilde{H}^{(i)} = H^{(i)} / \log_2 K$. Higher topic entropy indicates broader, more diverse topical coverage.

\textbf{Descriptive Ratio}~\citep{lu2010automatic} measures the proportion of adjectives (ADJ, if the token contains any of a list of [``ous'', ``ful'', ``able'', ``ible'', ``ive'', ``ish'', ``al'', ``ic'', ``ical'', ``like'', ``less'', ``y'', ``ary'', ``ory'', ``ant'', ``ent'']) and adverbs (ADV, if the token ends with ``ly'') among all alphabetic tokens, estimated via part-of-speech tagging (spaCy) when available, or by morphological heuristics otherwise:
\begin{equation}
    \text{Descriptive Ratio} = \frac{\sum_{w \in d} \mathbbm{1}[\text{POS}(w) \in \{\text{ADJ, ADV}\}]}{|d_{\text{alphabet}}|}
\end{equation}
where $|d_{\text{alphabet}}|$ is the count of alphabetic tokens. A higher value reflects more evaluative or elaborative language.

\textbf{Sentiment Compound}~\citep{hutto2014vader} captures the overall affective polarity of an instruction using the VADER lexicon, which returns a compound score $s \in [-1, 1]$, where $-1$ is maximally negative, $0$ is neutral, and $+1$ is maximally positive.

\textbf{Main IDF}~\citep{leskovec2020mining} reflects how lexically distinctive or specific a response is, based on the mean Inverse Document Frequency (IDF) of its top-$K$ TF-IDF-weighted terms:
\begin{equation}
    \text{Main IDF} = \frac{1}{K}\sum_{k=1}^{K} \text{IDF}(w_k), \quad \text{IDF}(w) = \log\frac{N+1}{n_w + 1} + 1
\end{equation}
where $N_I$ is the total number of instructions and $n_w$ is the number of instructions containing word $w$. Higher values indicate rarer, more distinctive vocabulary of instructions.

\textbf{Mean Content Length} measures the raw length of a response in terms of total token count, providing a simple proxy for verbosity:
\begin{equation}
    \text{Mean Content Length} = |d|
\end{equation}

To enable cross-metric comparison, all features were standardized across conditions using $z$-score normalization and subsequently mapped to $[0, 1]$ via the logistic function:
\begin{equation}
    \tilde{f} = \frac{1}{1 + \exp\!\left(-\frac{f - \mu_f}{\sigma_f}\right)}
\end{equation}
where $\mu_f$ and $\sigma_f$ are the mean and standard deviation of feature $f$ across all conditions. The normalized values were visualized as a radar plot (Fig. \ref{fig:figure3}D), with each axis representing one metric.

\section{Experimental Interface}\label{appendix:experiment-interface}

Here we present an example of the vibe coding experiment, which was also used as the teaching material in the experiment deployment.
\begin{figure}[H]
  \begin{center}
    \includegraphics[width=0.95\columnwidth]{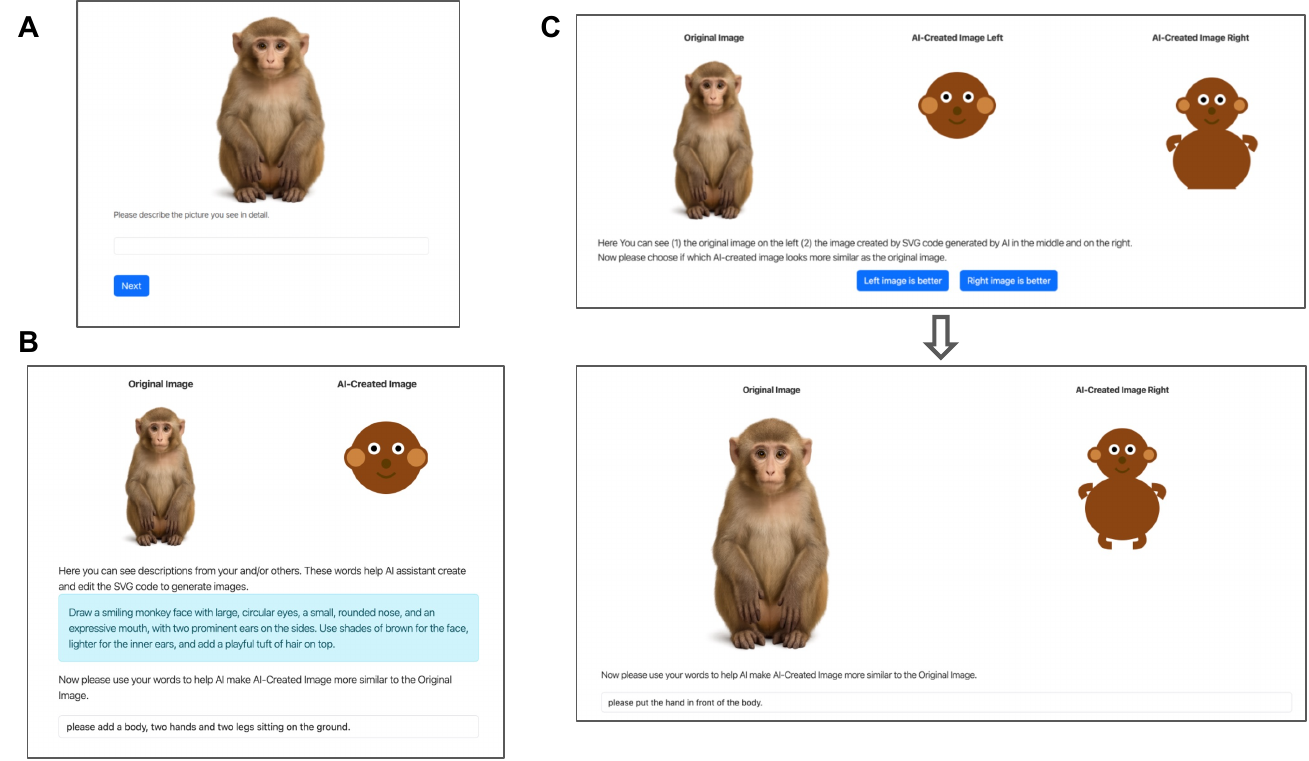}
  \end{center}
  \caption{Experimental interface. (A) Code Initialization. (B) Code editing as Fig. \ref{fig:figure1}A. (C) Code selection and editing as Fig. \ref{fig:figure1}B.}
  \label{fig:appendix-interface}
\end{figure}

\section{Validation Experiment Interface}\label{appendix:validation-interface}
\begin{figure}[H]
  \begin{center}
    \includegraphics[width=0.7\columnwidth]{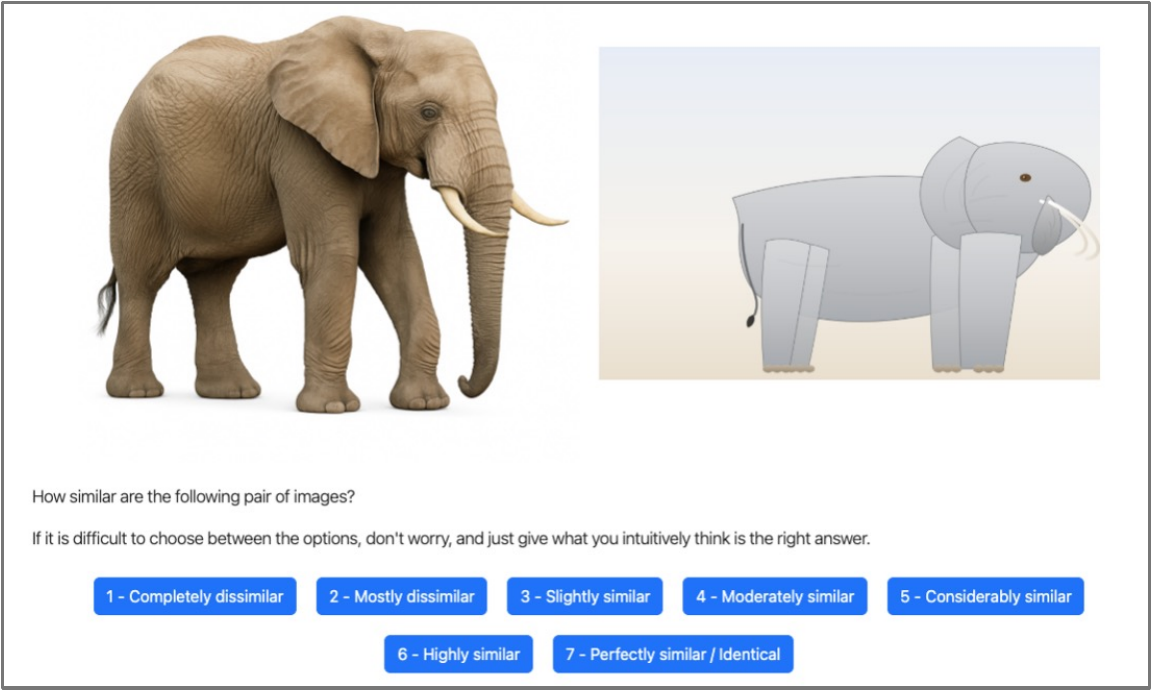}
  \end{center}
  \caption{Validation Rating Experiment Interface}
  \label{fig:appendix:interface-validation}
\end{figure}

\section{Validation Results with AI as the evaluator}\label{appendix:ai-rating}
\begin{figure}[H]
  \begin{center}
    \includegraphics[width=0.8\columnwidth]{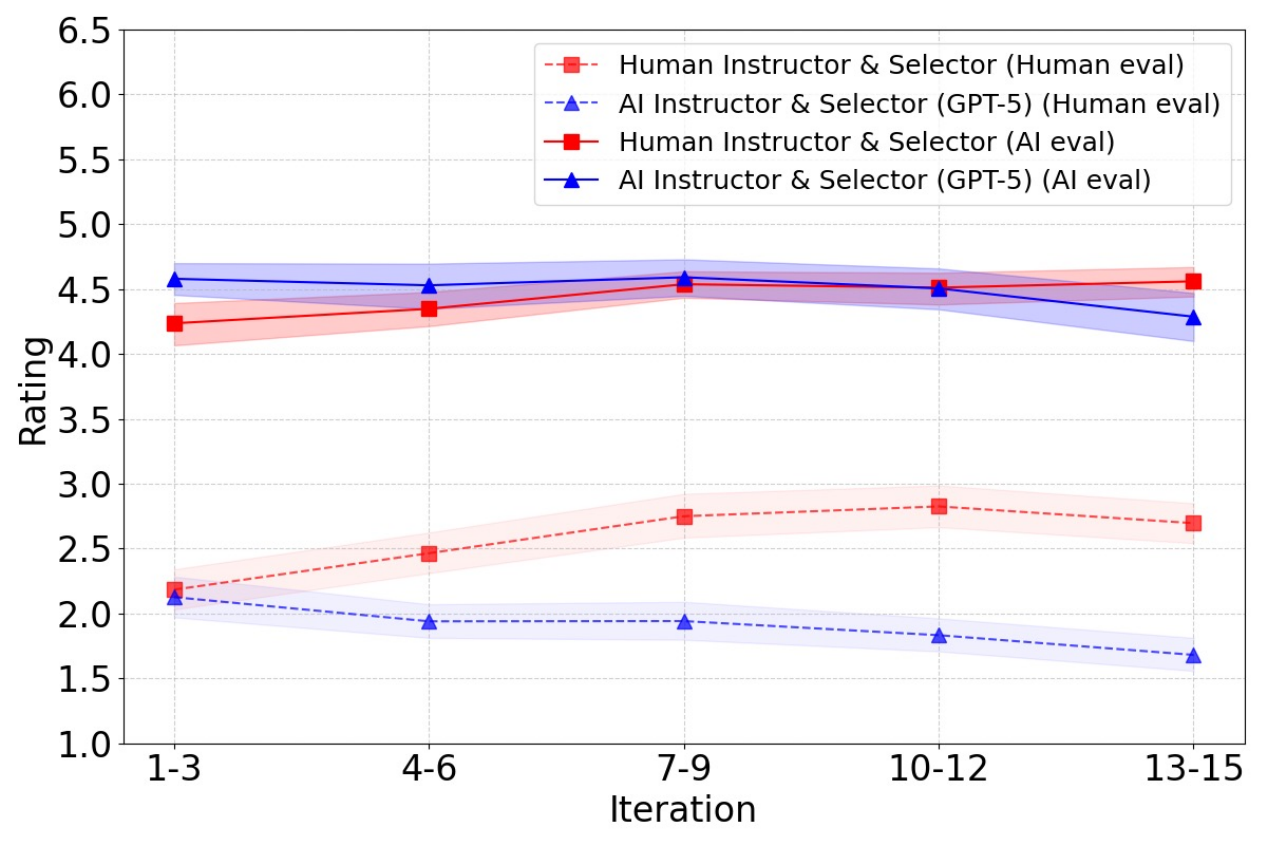}
  \end{center}
  \caption{Performance and difference with AI as evaluator compared with human as evaluator.}
  \label{fig:ai-rating}
\end{figure}

\section{Qualitative Analysis of Vibe Coding}\label{appendix:more-cases}
Here we provide more examples of vibe coding results extracted from the 15th iteration in the chain experiment. The first raw presents the reference images and the following presents different experimental conditions. In each subsection we all provide parts of examples from our two main human- and AI-led experiments as the baseline.

\subsection{SVG Examples of Semantic Controlled AI-Led Experiments}

\begin{figure}[H]
  \begin{center}
    \includegraphics[width=0.9\columnwidth]{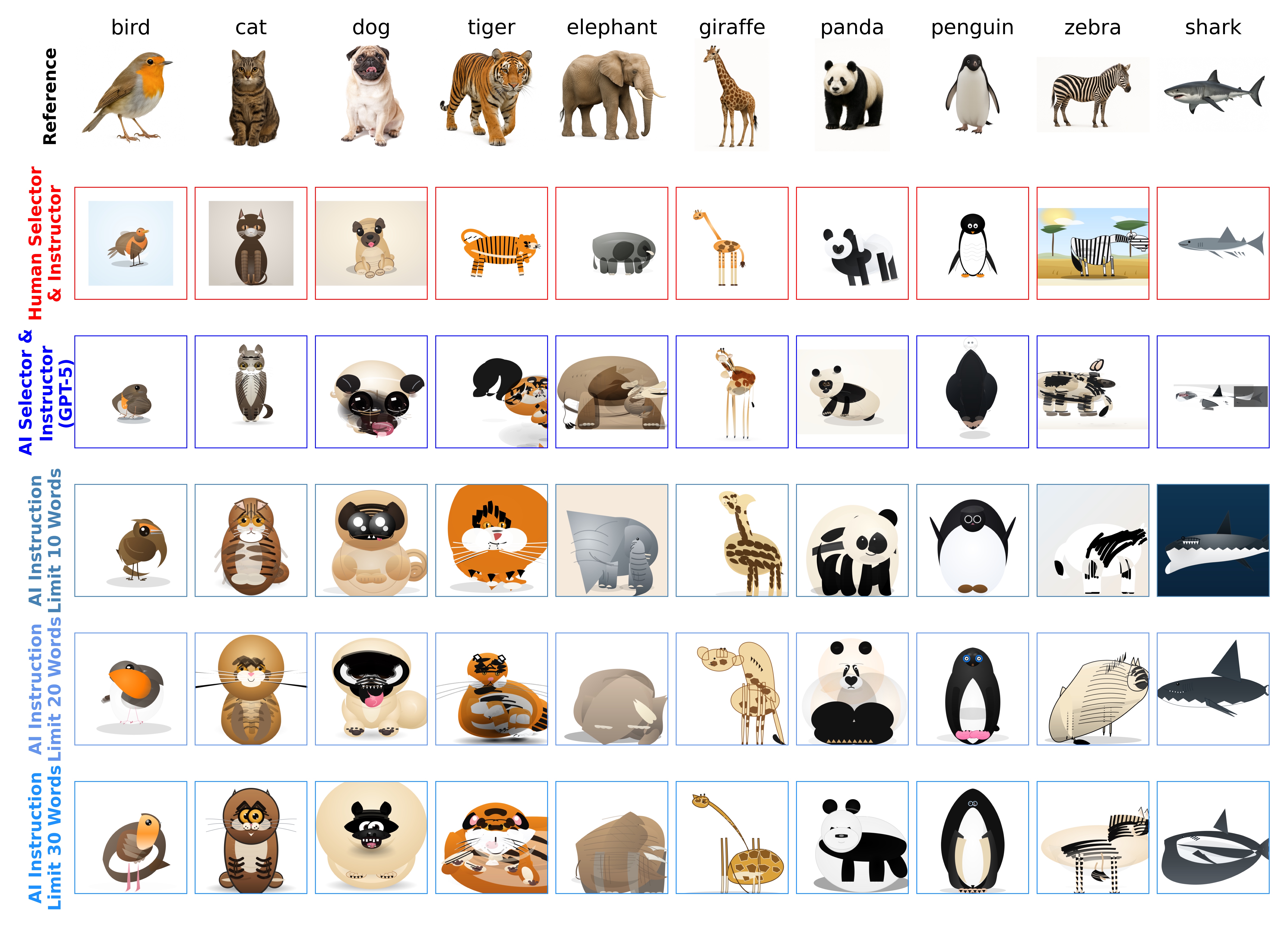}
  \end{center}
  \caption{Examples from all experiments  - Part 1.}
  \label{fig:grid-part1}
\end{figure}

\subsection{SVG Examples of Human-Led Asocial Experiment}
\begin{figure}[H]
  \begin{center}
    \includegraphics[width=0.9\columnwidth]{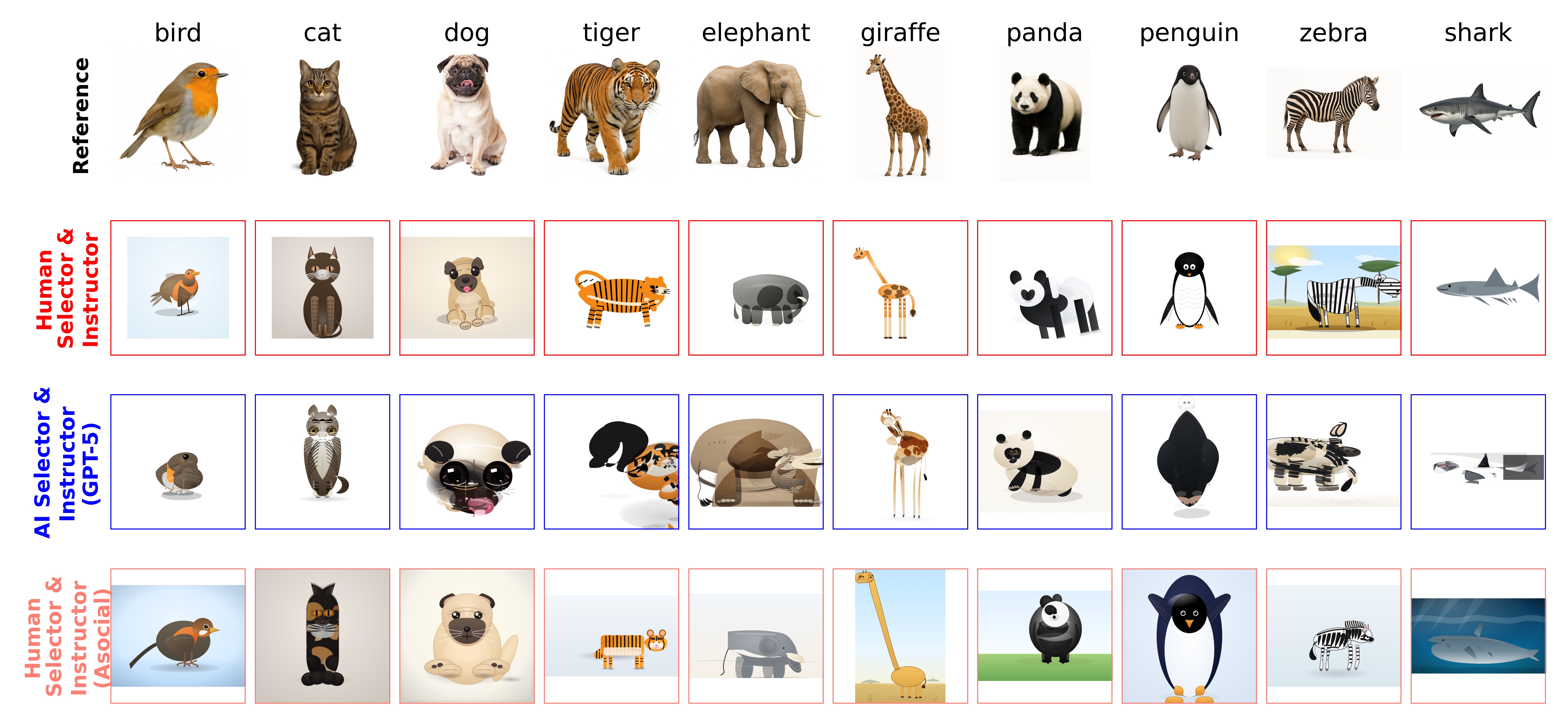}
  \end{center}
  \caption{Examples from all experiments  - Part 2.}
  \label{fig:grid-part2}
\end{figure}

\subsection{SVG Examples of Hybrid Human-AI Instructors \& Selectors}
\begin{figure}[H]
  \begin{center}
    \includegraphics[width=0.9\columnwidth]{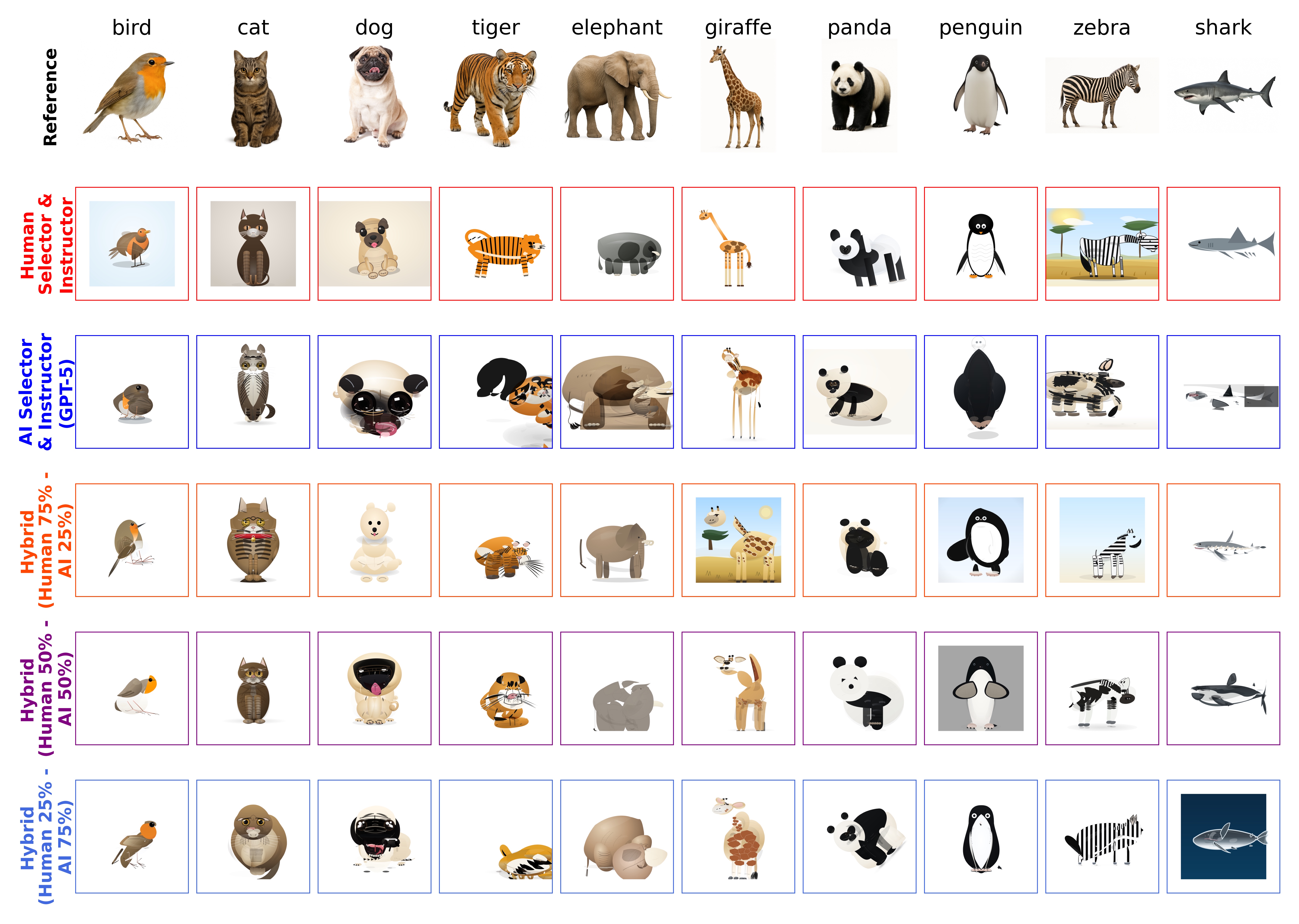}
  \end{center}
  \caption{Examples from all experiments  - Part 3.}
  \label{fig:grid-part3}
\end{figure}

\subsection{SVG Examples of Vibe Coding Role Division}
\begin{figure}[H]
  \begin{center}
    \includegraphics[width=0.9\columnwidth]{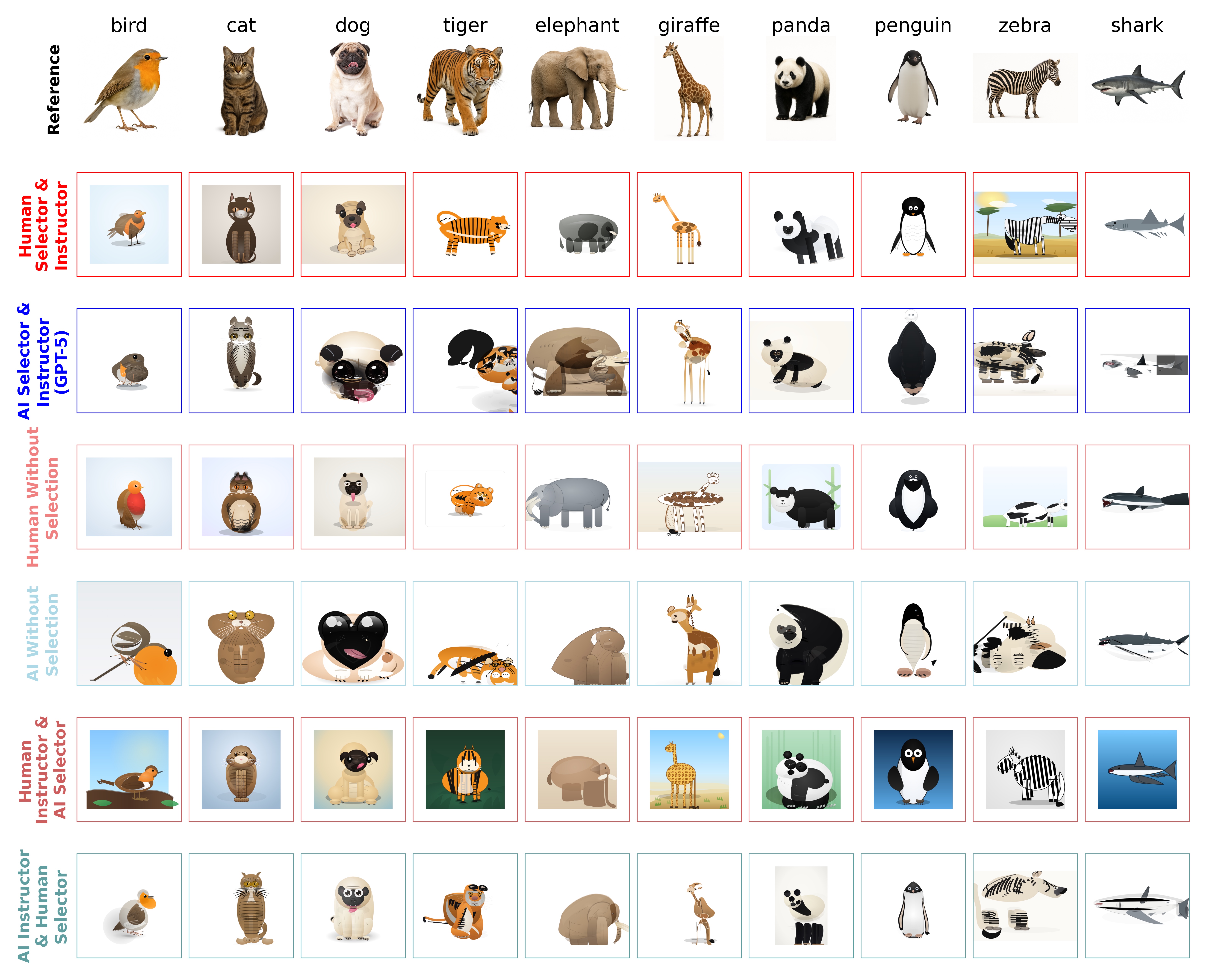}
  \end{center}
  \caption{Examples from all experiments  - Part 4.}
  \label{fig:grid-part4}
\end{figure}

\subsection{SVG Examples of AI Control Conditions}
\begin{figure}[H]
  \begin{center}
    \includegraphics[width=0.9\columnwidth]{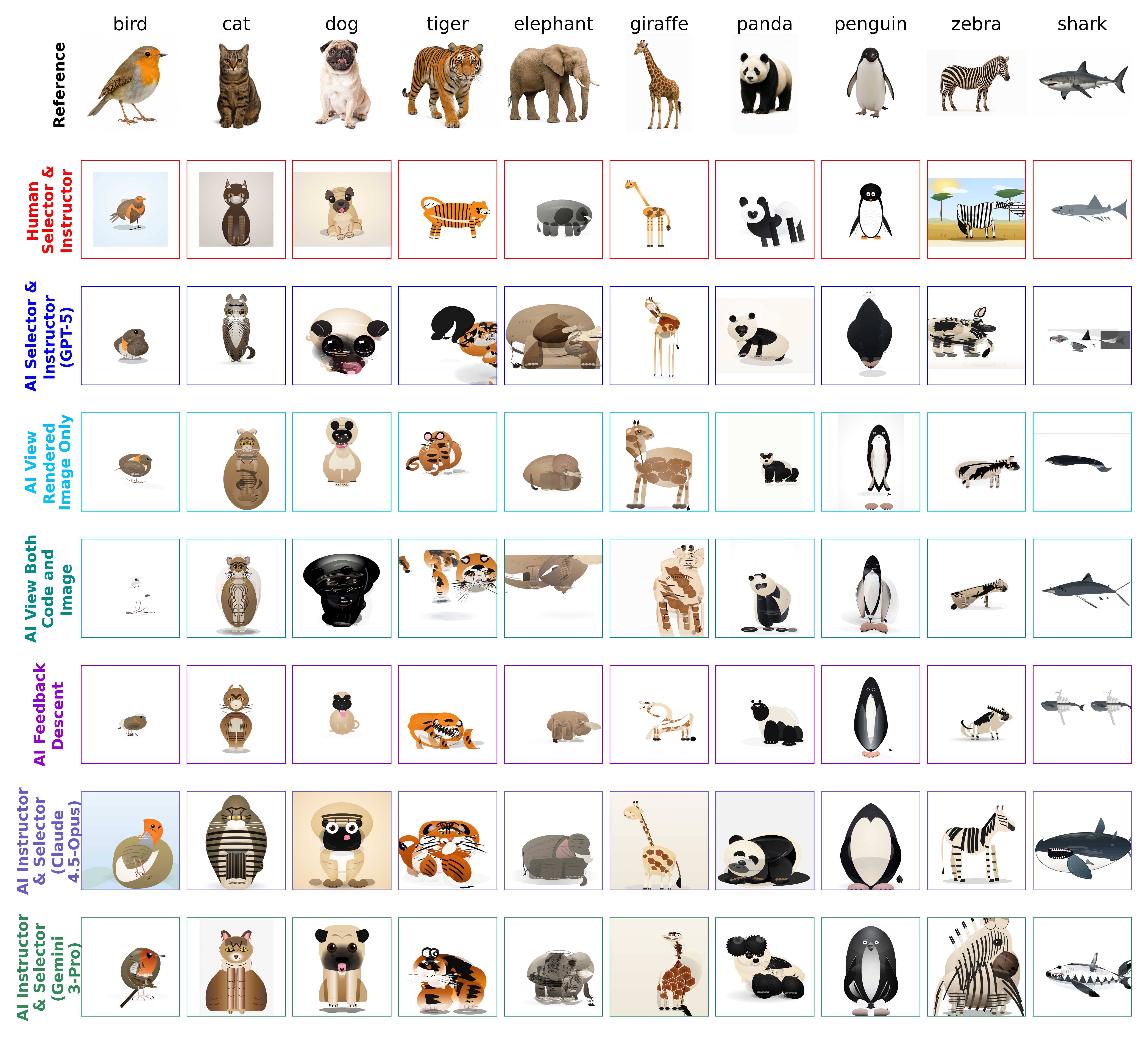}
  \end{center}
  \caption{Examples from all experiments  - Part 5.}
  \label{fig:grid-part5}
\end{figure}

\section{AI Prompts}\label{appendix:prompts}
\subsection{Code Generator}
\begin{lstlisting}
# SYSTEM ROLE
You are an AI assistant that creates and edits SVG (scalable vector graphics) based on user instructions.
Every time you create or edit an SVG, you should first fully understand the meaning of the instruction step by step.
With all instructions provided, you should understand the overall goal of the SVG code.
Then if any SVG code is provided, you should read it carefully and understand the current SVG code.
After that, you should create or edit the SVG code as detailed as possible according to both your understanding about the goal and the current instruction.
Your response must be ONLY raw SVG markup (no Markdown fences, no explanations).
Use a single <svg> root with xmlns='http://www.w3.org/2000/svg', width='512', height='512', and viewBox='0 0 512 512'.
Ensure the design fills the entire 512x512 canvas (no large white or empty areas)
Keep the SVG well-formed and executable, avoid <script>, <foreignObject>, <iframe>, or external image links.
Do not embed base64 images or any text labels unless explicitly requested.
When editing existing SVG, preserve useful parts and update only what is necessary to follow the new instruction.
\end{lstlisting}

If it is the first iteration (code initialization), we added the following part:
\begin{lstlisting}
## TASK
Create an SVG that matches this instruction: {instruction}.
Return ONLY the SVG markup, no explanations, no code fences.
\end{lstlisting}

Otherwise we added (pre\_instruction was provided only for iteration 2):
\begin{lstlisting}
## TASK
Edit the previous SVG to satisfy this instruction: {instruction}.
Here is the previous SVG code: {svg_code} and previous instruction: {pre_instruction}.
Return ONLY the updated SVG markup, no explanations, no code fences.
\end{lstlisting}

\subsection{AI as Instructor and Selector}\label{appendix:prompt-ai-instructor-selector}
When calling AI in the main-stream experiments, we provided it with the experiment background:
\begin{lstlisting}
## Background of the Experiment
In this experiment, you will work with AI assistants to create and edit SVG codes, which can be rendered as images. 
The AI assistant will generate and edit SVG codes based on your / other participants' / AI's descriptions. 
You will be assigned to one of the following parts: 
(1) you will see an image and describe it using your own words.
(2) you will see SVG code created by the AI assistant based on your and/or others' descriptions. 
(3) you will see two SVG codes created by the AI assistant based on your and/or others' descriptions. You need to choose which one is more similar to the original image.
(4) you will see the original image and the SVG code chosen by previous participants. You need to use your words to help the created SVG code as similar as the original image.
\end{lstlisting}

If AI was the instructor for iteration 1, we added:
\begin{lstlisting}
Please describe the picture you see in detail.
\end{lstlisting}

If AI was the instructor for iteration no less than 2 we added, the first sentence was only provided when iteration equaled 2:
\begin{lstlisting}
Here you can see descriptions from your and/or others. These words help AI assistant create and edit the SVG code:{instruction}.
Now you can see the original image:
{"url": f"data:image/png;base64,{rendered_base64_image}"}
{if viewing mode in ['view_both', 'view_code_only']}
here is the SVG code: \nCurrent SVG code:\n```svg\n{svg_preview}\n```
{if viewing mode in ['view_both', 'view_rendered_image_only']}
here is the rendered image:
{"url": f"data:image/png;base64,{rendered_base64_image}"}
\end{lstlisting}

If the instruction is limited to a certain length (10/20/30 words), we added:
\begin{lstlisting}
Limit your answer to be less than {output_limit} words.
\end{lstlisting}

If AI was the selector we added:
\begin{lstlisting}
Here You can see the original image and you will be provided with products created by AI as follows:
You will see the SVG code for each option.
SVG Code 1:\n```svg\n{svg_preview}\n```
SVG Code 2:\n```svg\n{svg_preview}\n```
Please choose which option is most similar to the original image.
If you think the first svg code is more similar, answer with 1, otherwise answer with 2.
Only provide the number as your answer, do not give any other texts.
\end{lstlisting}
This could be changed accordingly when the viewing mode was viewing rendered image only or viewing both.

If the experiment was Feedback Descent we added on the basis of the previous chunk except for the last two sentences:
\begin{lstlisting}
After choosing the best option, you must also provide:
1. Why the chosen option is better than the others
2. How the chosen option could be improved further
Please provide your response in the following format:
CHOICE: [number]
WHY_BETTER: [explanation of why this option is better]
HOW_TO_IMPROVE: [suggestions for how to improve the chosen option]
\end{lstlisting}

\subsection{AI as Evaluator}
\begin{lstlisting}
## System Role
You are an AI evaluator and your task is to evaluator how similar two images are and give a score.
## Evaluation Levels
You will have seven response options, ranging from 1 (Completely dissimilar) - 7 (Perfectly similar). Choose the one you think is most appropriate.
1 - Completely dissimilar
2 - Mostly dissimilar
3 - Slightly similar
4 - Moderately similar
5 - Considerably similar
6 - Highly similar
7 - Perfectly similar / Identical
## Evaluation Result
You should give your answer by choosing from one of these five levels and give the number.
You should only give one number as your evaluation result, please do not give any other texts in your response.
If it is difficult to choose between the options, don't worry, and just give what you intuitively think is the right answer.
## Task Instruction
Now you can see two images, please give your evaluation result only with number.
{"url": f"data:image/jpeg;base64,{encode_image(original_img_path)}}
{"url": f"data:image/jpeg;base64,{encode_image(test_img_path)}"}
\end{lstlisting}

\subsection{Rubric Generation for Feedback Descent}\label{appendix:prompt-rubric}
\begin{lstlisting}
## System Role
You are an expert art critic and evaluator. Your task is to create a detailed evaluation rubric for SVG illustrations based on an original reference image.

The rubric should follow this structure:
- RUBRIC NAME: [descriptive name]
- INTENT: [what matters most for this image]
- NON-NEGOTIABLES: [essential requirements]
- CRITICAL BENCHMARKS: [must evaluate these first with specific criteria]
- WHAT TO REWARD: [positive criteria to look for]
- WHAT TO PENALIZE: [negative criteria to avoid]
- TIEBREAKERS: [decision criteria when both are close]

Make the rubric specific, actionable, and operational - translate the aesthetic of the original image into concrete evaluation rules.

## Example Rubric

RUBRIC NAME: Anatomical Realism

INTENT: Believable equine anatomy with a plausible horn; form, proportion, and structure matter most.

NON-NEGOTIABLES:
- Recognizable equine proportions; head, neck, torso, four legs, mane, tail, horn present.
- Limbs connect anatomically; joints and hooves indicated.

CRITICAL BENCHMARKS (must evaluate these first):
1. Head-Neck Proportion: Neck length should be approx. 1.5x head length; head meets neck high on shoulders
2. Body Square: Body length (shoulder to buttock) = height at withers; chest depth = elbow height
3. Leg Structure: Proper joint articulation with elbow under withers; fetlock/pastern angles 45-55 deg when standing; all four limbs distinct and correctly connected

WHAT TO REWARD:
- Correct limb count and articulation; mass distribution that could stand or move.
- Horn integrates naturally with the skull (frontal bone center, 2-3" above eye line).
- Subtle shading or line variation conveying volume.
- Ground contact or cast shadow for grounding.
- Visible muscle definition suggesting tension/relaxation appropriate to pose.
- Differentiated hair textures: short coat vs coarse mane/tail strands.
- Anatomical landmarks: withers prominence, gaskin curve.

WHAT TO PENALIZE:
- Missing or fused legs; impossible joints; balloon torsos.
- Flat cardboard profiles with no sense of volume.
- Decorative effects that obscure structure.
- Disney-fied proportions (oversized eyes, baby-like features).
- Horn placement anywhere except frontal bone center (2-3" above eye line).

TIEBREAKERS:
- Prefer the image with more accurate limb/neck/head proportions.
- If both are plausible, choose the one with better weight and grounding.
\end{lstlisting}

When generating rubric for different reference images we added
\begin{lstlisting}
Analyze the following reference image and create a detailed evaluation rubric for SVG illustrations that should match this image.

The rubric should help evaluate:
1. How well an SVG illustration matches the original image
2. What specific aspects to look for (proportions, colors, shapes, details, etc.)
3. What makes a good match vs. a poor match

Create a rubric in the following format:

RUBRIC NAME: [name]
INTENT: [what matters most]
NON-NEGOTIABLES:
- [essential requirement 1]
- [essential requirement 2]

CRITICAL BENCHMARKS (must evaluate these first):
1. [benchmark 1 with specific criteria]
2. [benchmark 2 with specific criteria]
3. [benchmark 3 with specific criteria]

WHAT TO REWARD:
- [positive criterion 1]
- [positive criterion 2]

WHAT TO PENALIZE:
- [negative criterion 1]
- [negative criterion 2]

TIEBREAKERS:
- [tiebreaker 1]
- [tiebreaker 2]
\end{lstlisting}

\section{Examples of Rubric and Feedback}\label{appendix:rubric-feedback}
\subsection{Examples of Rubric}\label{appendix:example-rubric}
1. The rubric generated with cat as the reference image
\begin{lstlisting}
RUBRIC NAME: Front-Facing Mackerel Tabby Match (SVG)

INTENT: Capture the exact front-facing, seated silhouette, facial proportions, and mackerel tabby patterning of the reference cat in a clean, scalable vector.

NON-NEGOTIABLES:
- Full-body, front-facing seated cat on a white background with a soft, subtle ground shadow beneath the front paws.
- Warm brown base coat with darker near-black mackerel tabby stripes; amber irises with vertical pupils; pink-brown triangular nose; visible white whiskers.

CRITICAL BENCHMARKS (must evaluate these first):
1. Silhouette and Pose Fidelity
   - Upright seated posture; head centered; both front legs straight and parallel; both front paws fully visible and rounded.
   - Head height is approximately 35 to 40 percent of total figure height; body widest around chest/shoulders at approximately 1.3 to 1.6x the head width.
   - A small negative space gap between front paws (about 5 to 12 percent of body width); hindquarters implied symmetrically behind; tail not prominent from the front (may tuck along the right side without breaking symmetry).
2. Facial Layout and Proportions
   - Ears: two rounded triangles; ear height approximately 35 to 45 percent of head height; ear tips slightly splayed outward; inner ear lighter.
   - Eyes: large amber irises with vertical pupils; intereye spacing approx. one eye width; eye line sits roughly at the mid-to-slightly below mid point of the head (about 45 to 52 percent from chin to crown). Symmetrical circular catchlights in both eyes.
   - Nose and muzzle: upside down triangular pink-brown nose centered; nose width approximately 30 to 40 percent of intereye distance; lighter muzzle/chin patch with distinct whisker pads; multiple white whiskers radiating outward and slightly downward.
3. Tabby Pattern and Color Mapping
   - Clear "M" marking on the forehead between ears; dark cheek lines curving toward the whisker pads.
   - Narrow, vertically oriented mackerel stripes on neck and torso; banding on front legs; darker stripes over a warm brown base; central chest/neck fur visibly lighter.
   - Subtle dorsal to ventral shading: darker along spine/flanks, lighter underchest and inner legs; stripe edges clean and tapered, not blotchy or spotlike.

WHAT TO REWARD:
- Accurate, symmetrical face and ear/eye/nose relationships, with amber eye color, vertical pupils, and believable catchlights; whiskers rendered as fine light strokes that taper.
- Stripe layout that mirrors the reference: forehead M, cheek stripes, leg bands, and evenly spaced torso stripes with crisp, vector-clean edges and gentle tonal gradients suggesting fur volume.

WHAT TO PENALIZE:
- Wrong pose or silhouette (profile/three-quarter views, missing or fused paws, exaggerated tail presence from the front, head too large/small vs body, no paw gap).
- Incorrect pattern or palette (spots instead of narrow stripes, cool/gray overall hue, green/blue eyes, round pupils, missing forehead M, no chest lightening, no ground shadow or non-white background).

TIEBREAKERS:
- Prefer the SVG with more precise facial proportions (eye spacing, ear height/angle, nose size) and lifelike gaze.
- If both are close, choose the one with better tabby stripe placement and cleaner vector execution (smooth gradients, minimal banding, crisp edges at any scale).
\end{lstlisting}

2. The rubric generated with bird as the reference
\begin{lstlisting}
RUBRIC NAME: European Robin Side-Profile Fidelity

INTENT: Produce an SVG that convincingly matches the reference robin: right-facing side profile, crisp silhouette, correct color blocking (orange throat, brown wing, grey-buff body), believable leg/foot structure, and soft feather transitions on a clean white background.

NON NEGOTIABLES:
 Right facing European robin silhouette: compact round body, small rounded head, short straight beak, tail angled slightly downward and back, folded wing visible.
 Orange throat/upper chest patch present and correctly placed; black glossy eye with catchlight; two visible legs with three forward toes and one rear toe per visible foot.
 Clean white background; no extraneous props or textures.

CRITICAL BENCHMARKS (must evaluate these first):
1. Silhouette and Proportions
    Head to body: head height approx. 40 to 50 percent of body height; body an oval leaning slightly forward.
    Tail: length visible approx. 0.9 to 1.3x head height, angled 10 to 25 degrees downward from the body axis.
    Beak: short, straight, tapering; length approx. 2 to 2.5x eye diameter; alignment parallel to ground or with slight upward cant (<10 degree).
    Wing: folded along flank; tips stop before tail tip; top edge aligns with back curve.
2. Color Blocking and Placement
    Orange bib: saturated orange from throat to upper chest, feathered edge blending into pale belly; upper margin reaches under the eye but does not cross it.
    Back/wing: warm brown olive; belly/flanks light grey buff; subtle grey at crown.
    Eye: solid black with a small white highlight at approx. 10 to 11 o'clock; no colored iris.
3. Legs/Feet Structure and Grounding
    Two thin, scaly legs: front leg slightly forward, rear leg slightly back; joints kinked naturally.
    Each foot shows three forward toes and one rear hallux; claws slightly curved, darker tips.
    Optional soft contact shadow under feet; no floating.

WHAT TO REWARD:
 Accurate, clean silhouette with smooth Bezier curves; no kinks at neck, back, or belly.
 Faithful palette and value relationships:
   Orange bib: #F28B00 to #FFA62B range with soft gradient to belly.
   Wing/back browns: #7A5C3A to #8A7A5A; belly greys: #BDB8AF to #E4E0D8.
   Leg/feet pink-brown: #A8745A with darker claws #5A3B2C.
 Feather logic suggested via layered fills or clipped shapes (e.g., 3 to 5 layers on wing: coverts, secondaries, primaries) without heavy outlines.
 Light direction consistency (soft highlights from upper to left; belly and underwing slightly darker).
 Eye rendered as a crisp circle with a single specular highlight; subtle rim light acceptable.
 SVG craft: gradients used for bib to belly blend; minimal strokes (<=2 percent of canvas height), consistent stroke joins; scales cleanly at 64 to 512 px; no raster embeds.

WHAT TO PENALIZE:
 Wrong species cues: long/curved beak, elongated tail, mask patterns, or missing orange bib.
 Incorrect pose: facing left when the reference is right; tail horizontal or pointing up steeply; wing crossing past tail tip.
 Color errors: neon oranges, oversaturated browns, or flat single color fills with no value shift; orange patch intruding over the eye or too small.
 Eye mistakes: colored iris, no highlight, or oval/almond shape.
 Anatomical issues: fused legs, missing hallux, straight pipelike legs, or claws pointing the wrong way.
 Heavy black outlines, jagged curves, gradient banding, or visible anchorpoint kinks.
 Floating bird or hard drop shadows inconsistent with feet contact.

TIEBREAKERS:
 Choose the SVG with the more accurate silhouette and toe/claw anatomy.
 If silhouettes are close, prefer the one with better orange bib blending and eye highlight/lighting consistency.
\end{lstlisting}

\subsection{Examples of Feedback}\label{appendix:example-feedback}
Example 1
\begin{lstlisting}
  "why_better": "I can\u2019t evaluate\u2014both SVG renders failed to display. Selecting Option 1 as a placeholder until working previews or valid SVG files are provided.",
  "how_to_improve": "Please resend valid SVGs or screenshots. Ensure:\n Proper viewBox and explicit width/height so the image renders.\n White background and a soft ground shadow under the front paws.\n Front-facing seated silhouette with parallel front legs and a small gap between rounded paws.\n Warm brown base coat with narrow, near black mackerel stripes; clear forehead \u201cM\u201d, cheek lines, torso striping, and leg bands.\n Face: amber irises with vertical pupils and symmetrical catchlights; correctly spaced eyes; centered pink-brown triangular nose; lighter muzzle/chin; tapered white whiskers.\n Gentle dorsal to ventral shading and clean, tapered stripe edges for a vector clean look."
\end{lstlisting}

Example 2
\begin{lstlisting}
  "why_better": "I can\u2019t evaluate because neither SVG rendered, so a fair comparison isn\u2019t possible. Selecting Option 1 as a placeholder until working previews or SVG code are provided.",
  "how_to_improve": " Silhouette and proportions: right facing robin with compact oval body; head \u2248 40\u201350 percent of body height; short straight beak \u2248 2\u20132.5\u00d7 eye diameter; tail slightly down (10\u201325\u00b0), shorter than body and ending past wing tips.\n Color blocking: saturated orange throat/upper chest that reaches under the eye but does not cross it; warm brown olive back/wing; light grey-buff belly with soft gradient blend from the orange.\n Wing structure: suggest coverts/secondaries/primaries with 3\u20135 layered shapes; wing tips stop before tail tip; smooth B\u00e9zier curves along back and belly.\n Eye: solid black circular eye with a single small white highlight at approx. 10\u201311 o\u2019clock; no iris color.\n Legs/feet: two thin pink brown legs with natural kinks; each foot shows three forward toes and one rear hallux; slightly darker curved claws; subtle ground contact shadow.\n Craft: clean white background; minimal strokes; soft highlights from upper left; avoid heavy outlines or banded gradients; ensure scalable, no raster embeds."
\end{lstlisting}

\end{appendices}

\end{document}